\documentclass[amstex,thmsa,11pt,sw20bams]{article}%
\usepackage{amsmath}
\usepackage{amsfonts}
\usepackage[utf8]{inputenc}
\usepackage{graphicx}
\usepackage[francais]{babel}
\usepackage{amssymb}%
\providecommand{\U}[1]{\protect\rule{.1in}{.1in}}
\ifx\pdfoutput\relax\let\pdfoutput=\undefined\fi
\newcount\msipdfoutput
\ifx\pdfoutput\undefined\else
\ifcase\pdfoutput\else
\ifx\paperwidth\undefined\else
\ifdim\paperheight=0pt\relax\else\pdfpageheight\paperheight\fi
\ifdim\paperwidth=0pt\relax\else\pdfpagewidth\paperwidth\fi
\fi\fi\fi
\begin{document}

\author{Choulakian V. \& Allard J., Universit\'{e} de Moncton, Canada
\and vartan.choulakian@umoncton.ca, jacques.allard@umoncton.ca}
\title{Scale-Invariant Correspondence Analysis of Compositional Data }
\date{August 2025}
\maketitle

\begin{abstract}
Correspondence analysis is a dimension reduction technique for visualizing a
nonnegative matrix $\mathbf{N}=(n_{ij})$ of size $I\times J$, in particular
contingency tables or compositional datasets; but it depends on the row and
column marginals of $\mathbf{N}$. Three complementary transformations of the
data $T(\mathbf{N)}=(T(a_{i}n_{ij}b_{j}))$ render CA of $T(\mathbf{N)}$
invariant for any $a_{i}>0$ and $b_{j}>0$. First, Greenacre's scale-invariant
approach, valid for positive data. Second, Goodman's marginal-free
correspondence analysis, valid for positive or moderately sparse data. Third,
correspondence analysis of the sign-transformed matrix, sign$(\mathbf{N)}%
=(sign(n_{ij})),$ valid for sparse or extremely sparse data. We demonstrate
these three methods on four real-world datasets with varying levels of
sparsity to compare their exploratory performance.



\end{abstract}

\section{Overview of scale-invariant correspondence analysis}

Correspondence analysis (CA) and log-ratio analysis (LRA) are two popular
methods for the analysis and visualization of a two-way contingency table or
compositional dataset $\mathbf{N}=(n_{ij})$\textbf{ }of size $I\times J$ for
$i=1,...,I$ and $j=1,...,J$ and $n_{ij}\geq0.$

LRA is considered theoretically ideal because it is invariant to arbitrary
rescaling of rows and/or columns, making it fully scale-invariant.\ However,
it requires strictly positive cell data (nonsparse $n_{ij}>0$).

By contrast, CA can handle nonnegative cell data (sparse $n_{ij}\geq0$) but
lacks scale-invariance: its results depend on the marginal sums of rows and columns.

This paper develops a unified framework to render CA fully invariant under any
row and column scalings---what Goodman (1996) called \textquotedblleft
marginal-free\textquotedblright\ CA---to place it on equal footing with the
scale-invariant ideal of LRA. We present three data transformations
$T(\mathbf{N)}=(T(a_{i}n_{ij}b_{j}))$ that guarantee CA on $T(\mathbf{N})$ is
invariant for all positive scaling factors $a_{i}>0$ and $b_{j}>0.$ Two of
these transformations also convert the data into a bistochastic form.

\subsection{Three Transformations to achieve scale-invariant CA}

\textbf{1) Limiting power (Box--Cox) transformation}

Greenacre (2010, result 1) states that applying CA on $\mathbf{N}^{(\alpha
)}=(n_{ij}^{\alpha})$ for $\alpha\in(0,1]$ seamlessly transitions to uniformly
weighted LRA as $\alpha\rightarrow0$. Greenacre's approach is valid for
strictly positive data, see section 3.

\textbf{2) Sinkhorn (RAS-ipf) scaling}

Scale rows and columns by positive factors $a_{i}>0$ and $b_{j}>0$ and
iteratively adjust them to sum to one. This produces a unique bistochastic
matrix. Goodman (1996) sketched a marginal-free CA by applying CA (mfCA) to
this bistochastic table, accommodating positive to moderately sparse data, see
section 4.

Goodman (1996)'s mfCA aims to marry Pearson-based CA with Yule's ratio-based
LRA for nonnegative data. As stated above, for positive data Greenacre's
limiting power-transform approach fully satisfies Goodman's aim. Choulakian \&
Mahdi (2024) show mfCA is a special case of CA with pre-specified marginals,
tracing back to Benz\'{e}cri's work.

\textbf{3) Sign transformation}

Replace every nonzero entry by 1, that is, apply $T(\mathbf{N)}=(sign(a_{i}%
n_{ij}b_{j}))=(sign(n_{ij}))$. CA on this binary presence/absence table
handles sparse or extremely sparse data that defeat other methods. It remains
invariant to row and column rescalings but does not yield a bistochastic
matrix, see section 6.

\subsection{Bistochastic matrix}

A nonnegative matrix $\mathbf{D=}(d_{ij})$ for $i=1,...,I$ and $j=1,...,J,$ is
named bistochastic (or doubly stochastic) if the following two conditions are
satisfied: $\frac{1}{I}\sum_{i}d_{ij}=\frac{1}{J}\sum_{j}d_{ij}=1$ for
$i=1,...,I$ and $j=1,...,J;$ see among others, Marshall et al. (2009, chapter
2) and Loukaki (2023).\ It follows that $\sum_{i,j}d_{ij}=IJ.$ So
$\mathbf{Q=D/(}IJ\mathbf{)}$ is a probability measure with uniform marginals;
that is, $\frac{1}{J}=q_{+j}=\sum_{i}q_{ij}$ and $\frac{1}{I}=q_{i+}=\sum
_{j}q_{ij};$ Mosteller (1968) and Goodman (1996, Eq (39) and (40)) named
$\mathbf{Q}$ \textquotedblright the standardized
distribution\textquotedblright. So the concepts of doubly stochastic matrix
$\mathbf{D}$ and its associated probability measure $\mathbf{Q=D/(}%
IJ\mathbf{)}$ with uniform marginal weights are equivalent for nonnegative datasets.

According to Goodman (1996, section 6), the first step for mfCA is to
transform the contingency table $\mathbf{N=}(n_{ij})$ into a bistochastic
matrix \textbf{D} by row and column scalings, where $\mathbf{D}=(d_{ij}%
=a_{i}n_{ij}b_{j})$, $a_{i}>0$ and $b_{j}>0$ for $i=1,...,I$ and $j=1,...,J.$
Row and column scalings of a matrix consists of pre- and post-multiplying the
original matrix $\mathbf{N=\ }(n_{ij})$ by two diagonal matrices $Diag(a_{i})$
and $Diag(b_{j})$. The computational tool to achieve this will be the Sinkhorn
algorithm, also known as the RAS method or the iterative proportional fitting
(ipf) procedure; for a comprehensive survey see Idel (2016)\textbf{.}

\subsection{Contents}

In our opinion, the study of row and column scale-invariant CA of a dataset
\textbf{N} emerges in two distinct situations depending on the sparsity of
\textbf{N}:

a) $\mathbf{N}$ is \textit{positive, }that is nonsparse\textit{ (}$n_{ij}%
>0)$\textit{,} thus scalable to a unique bistochastic matrix by two different
transformations. First, by a limiting power (similar to Box-Cox)
transformation, which is equivalent to the uniformly weighted LRA association
index by Greenacre (2010, Result 1)'s Theorem. Second, by row and column
scalings by Sinkhorn (1964)'s algorithm implicitly suggested by Goodman (1996,
Eq (39)). Both Greenacre and Goodman approaches are related: Goodman approach
is a specific first-order approximation of Greenacre approach via uniformly
weighted LRA.

b) $\mathbf{N}$ is \textit{nonnegative (there is at least one zero cell).
}Here we can have two cases. Case 1)\textit{ }\textbf{N}\textit{\ }is scalable
to a unique bistochastic matrix\textit{ }\textbf{D}. This case is described by
Goodman (1996, Eq (39)) and CA of \textbf{D} is named mfCA. Case 2:
$\mathbf{N}$ is scalable to a matrix \textbf{D} having the structure of
multiple diagonal bistochastic blocks. We name CA of \textbf{D} multiple mfCA.
For extremely sparse data sets multiple mfCA is not valid; so we propose CA of
sign transformed data, known also as binary (0/1 or presence/absence) data.

This paper is organized as follows: Section 2 presents preliminaries
concerning CA and LRA association indices and their relationships with the
additive and multiplicative double centering; section 3 discusses Greenacre's
Theorem; section 4 presents Goodman's mfCA (cases 1 and 2); section 5 presents
Sinkhorn (RAS-ipf) algorithm; section 6 compares the three scale-invariant
transformations, plus the row transformation based on Benz\'{e}cri's
homogeneity principle; section 7 presents four case studies; and we conclude
in section 8.

Benz\'{e}cri (1973a, b) is the reference book on CA but difficult to read
because of his use of tensor notation. Among others, Greenacre (1984), Le Roux
and Rouanet (2004) and Murtagh (2005) are the best representatives of
Benz\'{e}cri's approach; Beh and Lombardo (2014) present a panoramic review of
CA. Taxicab CA (TCA) is a robust variant of CA which will be also used in this
paper; see in particular Choulakian (2006, 2016) and Choulakian, Allard and
Mahdi (2023).

\section{Preliminaries on the analysis of contingency tables}

We consider a two-way contingency table $\mathbf{N}=(n_{ij})$ for $i=1,...,I$
and $j=1,...,J$, and $\mathbf{P=N/}n=(p_{ij})$ of size $I\times J$ the
associated correspondence matrix (probability table) of the contingency table
\textbf{N}. We define as usual $p_{i+}=\sum_{j=1}^{J}p_{ij}$, $p_{+j}%
=\sum_{i=1}^{I}p_{ij},$ the vector $\mathbf{(}p_{i+})\in%
\mathbb{R}
^{I},$ the vector $\mathbf{(}p_{+j})\in%
\mathbb{R}
^{J}$, and $\mathbf{M}_{I}=Diag(p_{i+})$ the diagonal matrix having diagonal
elements $p_{i+},$ and similarly $\mathbf{M}_{J}=Diag(p_{+j}).$ We suppose
that $\mathbf{M}_{I}$ and $\mathbf{M}_{J}$ are positive definite metric
matrices of size $I\times I$ and $J\times J$, respectively; this means that
the diagonal elements of $\mathbf{M}_{I}$ and $\mathbf{M}_{J}$ are strictly positive.

First, we review the CA and LRA association indices, then double centering of
a dataset.

\subsection{Association indices}

The CA association index of a probability table \textbf{P} is
\begin{equation}
\Delta_{ij}(\mathbf{P})=\frac{p_{ij}}{p_{i+}p_{+j}}-1, \tag{1}%
\end{equation}
where $\frac{p_{ij}}{p_{i+}p_{+j}}$ is named the density function of $p_{ij}$
with respect to the product measure $p_{i+}p_{+j}.$

Assuming $p_{ij}>0,$ we define the LRA association index (named log-linear
association index by Goodman (1996) with uniform weights to be%

\begin{equation}
\lambda(p_{ij},w_{i}^{R}=1/I,w_{j}^{C}=1/J)=\log(p_{ij})+\frac{1}{IJ}%
\sum_{i,j}\log(p_{ij})-(\frac{1}{I}\sum_{i}\log(p_{ij})+\frac{1}{J}\sum
_{j}\log(p_{ij})). \tag{2}%
\end{equation}

Equation (2) appears also in compositional data analysis, see Aitchison
(1986), and in spectral mapping of Lewi (1976).

\subsection{Scale-invariance}

The CA association index (1) depends on the table's marginal distributions and
will shift if you change any row or column sum, whereas the LRA association
index (2) is unaffected by any positive rescaling of rows or columns, that we
describe. Let $\tau_{ij}(p_{ij})$ represent an interaction (association)
index, such as $\Delta_{ij}(\mathbf{P})$ or $\lambda(p_{ij},w_{i}%
^{R}=1/I,w_{j}^{C}=1/J)$ described in equations (1) and (2) respectively.
Following Yule (1912) and Goodman (1996), we state the following\bigskip

\textbf{Definition 1}: An interaction index $\tau_{ij}(p_{ij}=\frac{n_{ij}%
}{\sum_{i,j}n_{ij}})$ is scale-invariant if $\tau_{ij}(p_{ij})=\tau
_{ij}(q_{ij}=\frac{a_{i}n_{ij}b_{j}}{\sum_{i,j}a_{i}n_{ij}b_{j}})$ for
arbitrary positive scales $a_{i}>0$ and $b_{j}>0$.\bigskip

The CA index $=\frac{p_{ij}}{p_{i+}p_{+j}}-1$ fails this property. Lemma 1
generalizes
\[
\lambda(\frac{p_{ij}}{p_{i+}p_{+j}},w_{i}^{R}=p_{i+},w_{j}^{C}=p_{+j}%
)=\lambda(p_{ij},w_{i}^{R}=p_{i+},w_{j}^{C}=p_{+j})
\]
stated in Goodman (1996).\bigskip

\textbf{Lemma 1}

a) The LRA association index (3) with prespecified positive weights
($w_{i}^{R},w_{j}^{C})$
\begin{equation}
\lambda(p_{ij},w_{i}^{R},w_{j}^{C})=\log(p_{ij})+\sum_{i,j}w_{i}^{R}w_{j}%
^{C}\log(p_{ij})-(\sum_{i}w_{i}^{R}\log(p_{ij})+\sum_{j}w_{j}^{C}\log(p_{ij}))
\tag{3}%
\end{equation}
is scale-invariant. In particular the LRA association index (2) with uniform
weights is scale-invariant.

b) The first-order approximation of (3) is%
\[
\lambda(p_{ij},w_{i}^{R},w_{j}^{C})\approx\frac{p_{ij}}{w_{i}^{R}w_{j}^{C}%
}-\frac{p_{i+}}{w_{i}^{R}}-\frac{p_{+j}}{w_{j}^{C}}+1.
\]
For a proof see Choulakian, Allard and Mahdi (2023).\bigskip

\textbf{Remarks:}

a) A result stated in Cuadras, Cuadras, and Greenacre (2006) is
\[
\lambda(p_{ij},w_{i}^{R}=p_{i+},w_{j}^{C}=p_{+j})\approx\Delta_{ij}%
(\mathbf{P})=\frac{p_{ij}}{p_{i+}p_{+j}}-1.
\]

b) We emphasize the fact that in general
\[
\lambda(p_{ij},w_{i}^{R}=1/I,w_{j}^{C}=1/J)\approx IJp_{ij}-Ip_{i+}%
-Jp_{+j}+1,
\]
is \textit{not related} to CA association index $\Delta_{ij}(\mathbf{P})$,
unless $\mathbf{P}$ is bistochastic, and in this particular case Goodman's
mfCA association index is a specific first-order approximation of Greenacre's
scale-invariant CA association index via Lemma 1b, see subsection 4.1.

\subsection{Double centering}

Both the CA association index $\Delta_{ij}(\mathbf{P})$ and the LRA
association index $\lambda(p_{ij},w_{i}^{R}=1/I,w_{j}^{C}=1/J)$ are double
centered; that is:%
\begin{align}
0  &  =\sum_{i=1}^{I}p_{i+}\Delta_{ij}(\mathbf{P})=\sum_{i=1}^{I}\frac{1}%
{I}\lambda(p_{ij},w_{i}^{R}=1/I,w_{j}^{C}=1/J)\nonumber\\
&  =\sum_{j=1}^{J}p_{+j}\Delta_{ij}(\mathbf{P})=\sum_{j=1}^{J}\frac{1}%
{J}\lambda(p_{ij},w_{i}^{R}=1/I,w_{j}^{C}=1/J). \tag{4}%
\end{align}

According to Tukey (1977, chapter 10), there are two kinds of double centering
that he named \textit{row-PLUS-column} and \textit{row-TIMES-column}; we name
them \textit{additive} and \textit{multiplicative}, that we describe.

More generally, let $y_{ij}=h(p_{ij})$ be any transform of $p_{ij}$, and
define the three weighted means $Y_{i+}=\sum_{j=1}^{J}y_{ij}w_{j}^{C},$
$Y_{+j}=\sum_{i=1}^{I}y_{ij}w_{i}^{R}$ and $Y_{++}=\sum_{j=1}^{J}\sum
_{i=1}^{I}y_{ij}w_{i}^{R}w_{j}^{C}.$ Two centering schemes arise:

\textbf{a) Multiplicative} (row-times-column):
\begin{equation}
\tau_{ij}=y_{ij}-\frac{Y_{i+}Y_{+j}}{Y_{++}}, \tag{5}%
\end{equation}
where rank($\tau_{ij})=\ $rank$(y_{ij})-1.$

\textbf{b) Additive} (row-plus-column):
\begin{equation}
\tau_{ij}=y_{ij}+Y_{++}-(Y_{i+}+Y_{+j}), \tag{6}%
\end{equation}
where rank($\tau_{ij})=\ $rank$(y_{ij})-1$ or rank$(y_{ij})-2.$

Only the additive version is invariant to adding a constant $a_{i}$ to the
$i$th row or $b_{j}$ to the $j$th column of ($y_{ij})$: Thus Lemma 1 becomes a
corollary to this fact.

We consider two functional forms of $y_{ij}=h(p_{ij}).$

a) $y_{ij}=h(p_{ij})=\log p_{ij}$ for $p_{ij}>0.$ First, the log-linear
interaction $\tau_{ij}\mathbf{=}\lambda(p_{ij},w_{i}^{R},w_{j}^{C})$ is
obtained from (6). So the rank $\mathbf{(}\lambda(p_{ij},w_{i}^{R},w_{j}%
^{C}))=$ rank($\log p_{ij})-1,$ as in Goodman (1991, Table 11); or rank
$\mathbf{(}\lambda(p_{ij},w_{i}^{R},w_{j}^{C}))=\ $rank($\log p_{ij})-2,$ as
in Goodman (1991, Table 10). Second, to our knowledge the log-interaction
obtained from (5) has not been applied yet, most probably due to the fact that
it is not row and column scale-invariant.

b) $y_{ij}=h(p_{ij})=\frac{p_{ij}}{w_{i}^{R}w_{j}^{C}}$ is the density
function of the joint probability measure $p_{ij}$ with respect to the product
measure $w_{i}^{R}w_{j}^{C}.$ The CA interaction $(\tau_{ij}\mathbf{)=(}%
\Delta_{ij}(\mathbf{P}))$ is obtained either from (5) or from (6), when
($w_{i}^{R},w_{j}^{C})=(p_{i+},p_{+j}).$ So the rank$\mathbf{(}\Delta
_{ij}(\mathbf{P}))=\ $rank($\frac{p_{ij}}{p_{i+}p_{+j}})-1.$ This property of
CA seems \textit{unique}. So we ask the following question: Under what
condition the additive double centering equals the multiplicative double
centering, $(5)=(6)$? The proof of the next lemma is easy.\bigskip

\textbf{Lemma 2}: Assuming $y_{ij}\geq0$, equations $(5)=(6)=y_{ij}-Y_{++}$ if
and only if $Y_{++}=Y_{i+}=Y_{+j}$ for $i=1,...,I$ and $j=1,...,J$.\bigskip

\textbf{Corollary 1}:

a) In CA, taking $y_{ij}=\frac{p_{ij}}{p_{i+}p_{+j}}$ give all three marginals
equal to 1:
\[
Y_{i+}=\sum_{j}p_{+j}\frac{p_{ij}}{p_{i+}p_{+j}}=Y_{++}=\sum_{i,j}p_{i+}%
p_{+j}\frac{p_{ij}}{p_{i+}p_{+j}}=Y_{+j}=1;
\]
so both centering schemes produce $\Delta_{ij}(\mathbf{P})=\frac{p_{ij}%
}{p_{i+}p_{+j}}-1$; see also Goodman (1996, Eq (11)).

b) If $\mathbf{Y}$ is any bistochastic matrix, then it too has uniform
marginals, making the two centering formulas identical for $y_{ij}=IJp_{ij}$.

\subsection{Underlying structure via CA}

Following Benz\'{e}cri (1973b, p.6)'s famous second principle '\textit{Le
mod\`{e}le doit suivre les donn\'{e}es, non l'inverse}', \textquotedblright
the model must follow the data, not the inverse\textquotedblright, one can
read off the underlying structure of a data set from CA (or TCA) outputs such
as figures or dispersion values, see in particular Benz\'{e}cri (1973b,
pp.43-46 and pp. 188-195) and Benz\'{e}cri (1973a, pp. 261-287).

The following two results are well known for sparse datasets \textbf{N}:

\textbf{Theorem} (Benz\'{e}cri (1973b, pp188-190)) If the first
$m<rank(\mathbf{P})-1$ singular values of CA of $\mathbf{P}$ all equal 1,
$\rho_{\alpha}=1$ for $\alpha=1,...,m,$ then the latent underlying structure
of the contingency table \textbf{N} decomposes into ($m+1)$ independent
blocks, and CA reduces to separate analyses of each block.

\textbf{Remark 1}: Benz\'{e}cri (1973b, p.189-190) observed that it is rare to
have $\rho_{1}=1$, but not uncommon to have $\rho_{1}^{2}\geq0.7$ or $\rho
_{1}\geq0.837$; then the underlying structure of the contingency table may be
either quasi-2 blocks diagonal or band diagonal.

\subsection{Sparsity indices in CA}

Given $\mathbf{N}=(n_{ij})$ of size $I\times J$ three indices of sparsity (in
\% or proportion of zero cells $n_{ij}=0$) are defined within CA framework,
for the first two see Choulakian (2017):

a) \textbf{Apparent sparsity} of (\textbf{N}) in \%\ = 100$\frac{\text{number
of }(n_{ij}=0)}{IJ}.$

b) Compute \textbf{N}$_{merged}=(n_{ij}^{\ast})$ of size $I_{1}\times J_{1},$
where $I_{1}\leq I$ and $J_{1}\leq J,$ by combining rows and columns of
\textbf{N} which are proportional. Note that CA (or TCA) of \textbf{N} is
equivalent (identical) to CA (or TCA) of \textbf{N}$_{merged}$ based on
Benz\'{e}cri's principle of distributional equivalence. Then%

\begin{align*}
\text{\textbf{CA sparsity} of }(\mathbf{N})  &  =\text{Apparent sparsity of
}(\mathbf{N}_{merged})\\
&  =\frac{\text{number of }(n_{ij}^{\ast}=0)}{I_{1}J_{1}}.
\end{align*}

c) According to Loukaki (2023, Theorem 1) the sparsest nonnegative
bistochastic matrix $\mathbf{S}$ of size $I_{1}\times J_{1}$ has the smallest
support (number of positive cells) of $S(I_{1},J_{1})=I_{1}+J_{1}-\gcd
(I_{1},J_{1});$ so
\[
\text{sparsity of (}\mathbf{S)}=\frac{I_{1}J_{1}-S(I_{1},J_{1})}{I_{1}J_{1}}.
\]
Define
\[
\text{\textbf{Adjusted sparsity} of }(\mathbf{N})=\frac{\text{CA sparsity of
}(\mathbf{N})}{\text{sparsity of }(\mathbf{S}\text{ of size }I_{1}\times
J_{1})_{{}}}.
\]

Note that the sparsest bistochastic square matrices $\mathbf{S}$ are diagonal
with sparsity index of $(1-1/I_{1})$.

\textbf{Example 1}: Let
\[
\mathbf{N}=\left[
\begin{tabular}
[c]{ccccc}%
1 & 1 & 1 & 1 & 1\\
1 & 1 & 0 & 0 & 1\\
1 & 1 & 0 & 0 & 1\\
1 & 1 & 1 & 1 & 1\\
1 & 1 & 1 & 1 & 1\\
1 & 1 & 1 & 1 & 1
\end{tabular}
\ \ \ \ \ \right]  ;
\]
By Benz\'{e}cri's principle of distributional equivalence, it is CA equivalent
to
\[
\mathbf{N}_{merged}=\left[
\begin{tabular}
[c]{cc}%
12 & 8\\
6 & 0
\end{tabular}
\ \ \ \ \ \right]
\]
So: apparent sparsity of $(\mathbf{N})=4/30,$ CA sparsity of $(\mathbf{N}%
)=1/4,$ and adjusted sparsity of $(\mathbf{N})=1/2.$

\section{Greenacre's Theorem}

Greenacre's Theorem equates both CA and LRA\ association indices in a very
simple way (that is, equations $(1)=(2)$): Let $\mathbf{P}^{(\alpha)}%
=(p_{ij}^{(\alpha)}=\frac{n_{ij}^{\alpha}}{\sum_{i,j}n_{ij}^{\alpha}})$ for
$i=1,...,I$ and \ $j=1,...,J\ $be the correspondence table of the simple power
transformed data, $p_{i+}^{(\alpha)}=\sum_{j}p_{ij}^{(\alpha)}$ and
$p_{+j}^{(\alpha)}=\sum_{i}p_{ij}^{(\alpha)}$ for $\alpha>0.$ We consider the
CA association index
\begin{equation}
\Delta_{ij}(\mathbf{P}^{(\alpha)})=\frac{p_{ij}^{(\alpha)}-p_{i+}^{(\alpha
)}p_{+j}^{(\alpha)}}{p_{i+}^{(\alpha)}p_{+j}^{(\alpha)}}. \tag{7}%
\end{equation}
Then, Choulakian (2023) and Greenacre (2024) proved the following result:

\textbf{Greenacre's Theorem }(Greenacre (2010, Result 1)

Under the assumption $n_{ij}>0,$ for $\alpha\rightarrow0$ and $\alpha>0$%
\begin{align}
\lim_{\alpha\rightarrow0}\frac{\Delta_{ij}(\mathbf{P}^{(\alpha)})}{\alpha}  &
=\lim_{\alpha\rightarrow0}(\frac{p_{ij}^{(\alpha)}}{p_{i+}^{(\alpha)}%
p_{+j}^{(\alpha)}}-1)/\alpha\nonumber\\
&  =\lambda(p_{ij},w_{i}^{R}=1/I,w_{j}^{C}=1/J),\text{ and by Lemma
1a}\nonumber\\
&  =\lambda(q_{ij}=a_{i}p_{ij}b_{j},w_{i}^{R}=1/I,w_{j}^{C}=1/J)\text{ for
}a_{i},b_{j}>0\tag{8}\\
&  =\lim_{\alpha\rightarrow0}\frac{\Delta_{ij}(\mathbf{Q}^{(\alpha)})}{\alpha
}.\nonumber
\end{align}

Moreover, the row and column margins of both $\mathbf{P}^{(\alpha)}$ and
$\mathbf{Q}^{(\alpha)}$ become uniform:
\begin{equation}
\lim_{\alpha\rightarrow0}p_{i+}^{(\alpha)}=\lim_{\alpha\rightarrow0}%
q_{i+}^{(\alpha)}=1/I\text{\ \ for }i=1,...,I\text{\ \ and\ \ }\lim
_{\alpha\rightarrow0}p_{+j}^{(\alpha)}=\lim_{\alpha\rightarrow0}%
q_{+j}^{(\alpha)}=1/J\ \text{for}\ j=1,...,J. \tag{9}%
\end{equation}
\bigskip

\textbf{Remark 2}: There are two important aspects in Greenacre's Theorem that
we want to emphasize.

a) It shows the conditions under which CA becomes row and column
scale-invariant, equation (8).

b) It implies that the row and column marginals of $\mathbf{P}^{(\alpha)}$ for
$\alpha>0$ and $\alpha\simeq0$ are almost uniform, equation (9); so
$\mathbf{P}^{(\alpha)}$ is bistochastic.

\subsection{CA decomposition}

The limit association (8) can be decomposed as%

\begin{align*}
\lambda(p_{ij},w_{i}^{R}  &  =1/I,w_{j}^{C}=1/J)=\lim_{\alpha\rightarrow
0}\frac{\Delta_{ij}(\mathbf{P}^{(\alpha)})}{\alpha}\\
&  =\sum_{m=1}^{M}\mu_{im}\nu_{jm}\rho_{m},
\end{align*}
where $M=rank(\mathbf{P})-1$ and the parameters ($\mu_{im},\nu_{jm},\rho_{m})$
satisfy, for $s,m=1,...,M$
\[
1=\sum_{i=1}^{I}\mu_{im}^{2}/I=\sum_{j=1}^{J}\upsilon_{jm}^{2}/J
\]%
\[
0=\sum_{i=1}^{I}\mu_{im}/I=\sum_{j=1}^{J}\upsilon_{jm}/J
\]%
\[
0=\sum_{i=1}^{I}\mu_{im}\mu_{is}/I=\sum_{j=1}^{J}\upsilon_{jm}\upsilon
_{js}/J\text{\ \ for }m\neq s.
\]

The parameter $\rho_{m}\in(0,1]$ admits two interpretations:

First, $\rho_{m}=corr(\mu_{mi},\nu_{mj})=\sum_{i,j}$ $\mu_{mi}\nu_{mj}p_{ij}$
is the Pearson correlation coefficient of the $m$th principal dimension row
and column scores $(\mu_{mi},\nu_{mj})$ within the framework of CA of
$\mathbf{P}^{(\alpha)}$.

Second, it represents the uniformly weighted intrinsic association within the
framework of Goodman (1991, Eq. 3.7)'s RC association model, which can be
expressed as a function of log-odds ratio%
\begin{align*}
\log(\frac{p_{ij}p_{i_{1}j_{1}}}{p_{ij_{1}}p_{i_{1}j}}  &  =\sum_{m=1}^{M}%
(\mu_{im}-\mu_{i_{1}m})(\upsilon_{jm}-\upsilon_{j_{1}m})\rho_{m}\\
&  =\lim_{\alpha\rightarrow0}\frac{\Delta_{ij}(\mathbf{P}^{(\alpha)}%
)+\Delta_{i_{1}j_{1}}(\mathbf{P}^{(\alpha)})-\Delta_{i_{1}j}(\mathbf{P}%
^{(\alpha)})-\Delta_{ij_{1}}(\mathbf{P}^{(\alpha)}}{\alpha}.
\end{align*}
The last relation is called tetra differences by Rao (1973, p.12); and it
perfectly reconciles Pearson correlation measure $\rho_{m}$ with Yule's
log-odds ratio measure, an effort attempted by Mosteller (1968) and by Goodman
(1996, section 10).

\section{Goodman's mfCA and Sinkhorn scaling}

In Greenacre's Theorem, the strict positivity assumption ($p_{ij}>0)$ is
essential and there is only one parameter $\alpha.$ Here, we relax that
assumption to nonnegativity ($p_{ij}\geq0$) and introduce $(I+J)$ scaling
parameters to transform the matrix $\mathbf{P}=(p_{ij})$ into a nonnegative
bistochastic matrix $\mathbf{Q}=(q_{ij}=a_{i}p_{ij}b_{j})$ using Sinkhorn
(1964)'s algorithm---also known as the RAS algorithm or the iterative
proportional fitting (ipf) procedure. Depending on the arrangement of zeros in
$\mathbf{P}$, two distinct cases arise, characterized by Sinkhorn and Knopp
(1967) and independently by Brualdi et al. (1966) as \textit{fully
indecomposable} and \textit{reducible}. As we will see in the applications,
the sparsity pattern of the original data plays a fundamental role.

\subsection{Case 1: Nonsparse or moderately sparse fully indecomposable
matrices}

When the dataset $\mathbf{N}$ is either strictly positive (nonsparse
$n_{ij}>0$) or nonnegative with a fully indecomposable structure (i.e., there
exists a perfect matching along a permutation of rows and columns), there is a
unique choice of positive scaling elements $a_{i},b_{j},\varphi_{i},\psi
_{j}>0$ such that

$\mathbf{Q=(}q_{ij})=(a_{i}p_{ij}b_{j})=(\varphi_{i}n_{ij}\psi_{j})$ is unique
and bistochastic:
\begin{equation}
\mathbf{Q}=(q_{ij}=a_{i}p_{ij}b_{j})\text{ \ such that }q_{i+}=1/I\text{ and
}q_{+j}=1/J\text{ } \tag{10}%
\end{equation}

Sinkhorn (1964) established this result for positive matrices, and Sinkhorn \&
Knopp (1967) along with Brualdi et al. (1966) extended it to nonnegative,
fully indecomposable matrices (with further examples for rectangular arrays
given by Loukaki (2023)).

For the resulting bistochastic matrix $\mathbf{Q}$, Goodman's mfCA association
index is defined as%

\begin{equation}
\Delta_{ij}(\mathbf{Q})=IJq_{ij}-1.\text{ } \tag{11}%
\end{equation}

\textbf{Remark:} For strictly positive (nonsparse $n_{ij}>0$) data, Goodman's
mfCA association index (11) is a first-order approximation of Greenacre's
scale-invariant CA association index
\begin{align*}
\lim_{\alpha\rightarrow0}\frac{\Delta_{ij}(\mathbf{P}^{(\alpha)})}{\alpha} &
=\lambda(p_{ij},w_{i}^{R}=1/I,w_{j}^{C}=1/J),\text{ by Lemma 1a}\\
&  =\lambda(q_{ij}=a_{i}p_{ij}b_{j},w_{i}^{R}=1/I=q_{i+},w_{j}^{C}%
=1/J=q_{+j})\text{ for }a_{i},b_{j}>0\\
&  \approx\Delta_{ij}(\mathbf{Q})=IJq_{ij}-1\text{ by Lemma 1b.}%
\end{align*}

\subsection{Case 2: Sparse or extremely sparse reducible matrices}

When $\mathbf{P}$ (or $\mathbf{N}$) exhibits a high degree of sparsity and its
nonzero entries cluster into k disjoint blocks, the Sinkhorn (RAS-ipf)
procedure yields block-wise scaling rather than a unique global solution. In
this reducible case, one obtains scaling elements $a_{i\beta}^{\ast}%
,b_{j\beta}^{\ast}>0$ for each block $\beta=1,...,k$, producing a
block-diagonal matrix%

\begin{align}
\mathbf{Q}^{\ast}  &  =(q_{\beta(ij)}^{\ast}=a_{i\beta}^{\ast}p_{ij}b_{j\beta
}^{\ast})\nonumber\\
&  =Diag(\mathbf{B}_{1}...\mathbf{B}_{\beta}...\mathbf{B}_{k})^{\prime},
\tag{12}%
\end{align}
where each submatrix $\mathbf{B}_{\beta}$ is itself bistochastic but may have
different uniform marginals. By Benz\'{e}cri's Theorem (see subsection 2.3),
CA of $\mathbf{Q}^{\ast}$ is equivalent to $k$ CA of $\mathbf{B}_{\beta}$ for
$\beta=1,...,k.$

The corresponding block-specific mfCA index takes the form%

\begin{equation}
\Delta_{ij}(\mathbf{P},a_{i\beta}^{\ast},b_{j\beta}^{\ast})=\frac{a_{i\beta
}^{\ast}p_{ij}b_{j\beta}^{\ast}}{c_{\beta}r_{\beta}}-1,\text{ } \tag{13}%
\end{equation}
where the terms $c_{\beta},r_{\beta}>0$ represent different uniform marginals
of the blocks $\beta=1,...k.$

\textbf{Example 2}: The following simple 2 blocks diagonal dataset is sparse
reducible%
\begin{align*}
\mathbf{N}  &  =\left[
\begin{tabular}
[c]{cc}%
\textbf{B}$_{1}$ & 0\\
\textbf{0} & \textbf{B}$_{2}$%
\end{tabular}
\ \ \ \ \ \ \ \ \ \ \right] \\
&  =\left[
\begin{tabular}
[c]{cccc}%
1 & 1 & 1 & 0\\
0 & 0 & 0 & 1\\
0 & 0 & 0 & 1
\end{tabular}
\ \ \ \ \ \ \ \ \ \ \right]  .
\end{align*}
Each of the 2 blocks \textbf{B}$_{1}=(1\ 1\ 1)$ and \textbf{B}$_{2}^{\prime
}=(1\ 1)$ is bistochastic. However, because there is no way to mix entries
across the two blocks, the overall matrix cannot be made globally
bistochastic. Instead, we treat it as two separate bistochastic blocks under
the reducible case.

\section{The Sinkhorn (RAS-ipf) algorithm}

The Sinkhorn algorithm (also known as the RAS or iterative proportional
fitting procedure) is a simple method for adjusting a contingency table so
that its row and column margins match prespecified targets. Because the
literature on this topic is extensive, we refer readers to Idel (2016) for a
comprehensive survey.

As Madre (1980) explains, there are three variants of this algorithm:

- Two variants alternate between scaling the row sums and scaling the column
sums (this alternating-update approach is the standard and most widely
studied; see, for example, Landa et al. (2022), Algorithm 2).

- The third variant updates both row and column marginals simultaneously.

In this paper---and in the R code provided below---we adopt the
simultaneous-update variant.

To assess convergence, we employ two criteria that a bistochastic matrix
$\mathbf{D}=(d_{ij})$ must satisfy (see Subsection 1.2 and Corollary 1b).

1) \textbf{C2dist}, which measures the deviation of the average row and column
sums from their target of 1:%

\[
C2_{dist}=\sum_{i,j}|(\frac{1}{I}\sum_{i}d_{ij}+\frac{1}{J}\sum_{j}%
d_{ij})-2|,
\]

2) \textbf{ratio}, which compares the total mass of $\mathbf{D}$ to the ideal
mass $IJ$:%

\[
ratio=\sum_{i,j}d_{ij}/IJ.
\]

If $\mathbf{D}$ is bistochastic, then $C2_{dist}=0$ and $ratio=1.$ The
following R code is used in the applications.$\bigskip$

\# alg3 Madre Simultaneous adjustments

\ \ \ nRow
$<$%
- nrow(dataMatrix)

\ \ \ nCol
$<$%
- ncol(dataMatrix)

\ \ \ rowPrUnif
$<$%
- rep(1/nRow, nRow)

\ \ colPrUnif
$<$%
- rep(1/nCol, nCol)

\ \ nit
$<$%
- 500 \ \ \ \ \ \ \# nit = number of iterations

\ \ qij
$<$%
- dataMatrix/sum(dataMatrix)

for (jj in 1:nit) \{

rowPROBA
$<$%
- apply(qij,1,sum)

colPROBA
$<$%
- apply(qij,2,sum)

dij
$<$%
- qij/(rowPROBA \%*\% t(colPROBA))

qij
$<$%
- dij/sum(dij)

Gj
$<$%
- t(rowPrUnif) \%*\% dij

Gi
$<$%
- dij \%*\% colPrUnif

C2
$<$%
- ((rep(1, nRow) \%*\% Gj) +(Gi \%*\% t(rep(1, nCol))))-2

C2dist = sum(abs(C2))

ratio = sum(dij)/(nRow*nCol)

su = c(jj,C2dist,ratio)

print(su)

\}

\section{Data transformations}

The main instigation of this paper is the bistochastic similarity of equations
(9) and (10). In (9) there is one parameter $\alpha\approx0$, while in (10)
there are $(I+J)$ positive scaling parameters $(a_{i},$ $b_{j}).$ To
understand the difference between (9) and (10), we introduce the concept of a
transformation that does not change the internal order (cellwise) structure in
a contingency table, and show that row and column scalings (Cases 1 and 2) in
mfCA are transformations that may change dramatically the cellwise structure
thus uncovering the local-micro level interactions among some rows and columns
in mfCA.\bigskip

\textbf{Definition 2}: A transformation $T$ of a contingency table
$\mathbf{N}=(n_{ij})$ keeps the \textit{internal order structure} unchanged if
it satisfies the following inequality:
\begin{equation}
n_{ij}\leq n_{i_{1}j_{1}}\ \text{then}\ \ T(n_{ij})\leq T(n_{i_{1}j_{1}}).
\tag{14}%
\end{equation}
$\bigskip$

It is evident that the simple power transformation by $T(n_{ij})=n_{ij}%
^{\alpha}$ for $\alpha>0$ in Greenacre's Theorem satisfies (14); while the
transformations in Cases 1 and 2 represented in equations (10) and (12) by
$T(p_{ij})=a_{i}p_{ij}b_{j}$ and $T(p_{ij})=a_{i\beta}^{\ast}p_{ij}b_{j\beta
}^{\ast}$ in general do not satisfy (14), as shown by the following simple
contrived example.\bigskip

\textbf{Example 3 and its interpretation:} We apply the Sinkhorn (RAS-ipf)
algorithm to the following table, a variant first studied by Sinkhorn(1964),
\[
\mathbf{N}=%
\genfrac{[}{]}{0pt}{}{0\ \ \ \ \ 2}{1\ \ \ \ \ 30}%
.
\]
After $nit=500$ iterations we get: $C2_{dist}=4.128788e-03\simeq0$,
$ratio=9.999990e-01\simeq1$ and
\begin{align*}
T(\mathbf{N)} &  \mathbf{=D}=%
\genfrac{[}{]}{0pt}{}{0\ \ \ \ \ \ \ \ \ \ \ \ \ \ \ \ \ \ \ \ \ \ \ \ \ \ \ \ 1.9980667962}{1.997934\ \ \ \ \ \epsilon
_{nit=500}=0.003995601}%
\\
&  \simeq%
\genfrac{[}{]}{0pt}{}{0\ \ \ \ \ \ \ \ \ \ \ \ \ \ \ \ \ \ \ \ \ \ \ \ \ \ \ \ 2}{2\ \ \ \ \ \ \ \ \ \ \ \ \ \ \ \ \ \ \ \ \ \ \ \ \ \ \ \ \ 0}%
\end{align*}
which does not satisfy (14) at 2 cells.

Let us interpret both tables $\mathbf{N}$ and $T(\mathbf{N)}$. In $\mathbf{N}%
$, row 2 and column 2 are intensely related for $n_{22}=30$, thus exhibiting a
global-macro level probability association; while in $T(\mathbf{N)}$ we
observe two local-micro level probability associations (row 2 and column 1)
and (row 1 and column 2) which are independent.

\subsection{Sign transformation}

Define the sign transformation $T(n_{ij})=sign(n_{ij}),$ which is invariant
under any positive row- and column-scalings
\[
sign(n_{ij})=sign(a_{i}n_{ij}b_{j})\text{ \ \ \ \ \ \ \ for \ }a_{i}>0\text{
and\ }b_{j}>0.
\]
This transformation satisfies property (14) and encodes a uniform association
between row $i$ and column $j$:

- $sign(n_{ij})=1$ indicates presence of association,

- $sign(n_{ij})=0$ indicates absence of association.

This approach is especially valuable for sparse or extremely sparse matrices.

Moreover, recall that for any vector $\mathbf{x}\in R^{p}$, the $l_{0}$-norm
(which is not a norm!) $\parallel\mathbf{x}\parallel_{0}=\lim_{\alpha
\rightarrow0}\sum_{j}|x_{j}|^{\alpha}$ counts the number of nonzero entries in
\textbf{x}. Thus, when we perform CA on $sign(\mathbf{N})$, the weight
assigned to each row or column equals the $l_{0}$-norm of that row or column.

The small data set above in Example 3 becomes%
\[
sign(\mathbf{N)}=%
\genfrac{[}{]}{0pt}{}{0\ \ \ \ \ 1}{1\ \ \ \ \ 1}%
,
\]
where all positive cellwise probabilities are equal.\bigskip

\textbf{Remarks}: 

a) For nonegative compositional datasets, power transformation $T(n_{ij}%
)=n_{ij}^{\alpha}$ for $\alpha\in\lbrack0,1]$ is termed $\alpha$%
-transformation by Tsagris et al. (2011, 2023) and chiPower transformation by
Greenacre (2024). Their approach is row stochatic (RS): $p_{i+}^{(\alpha
)}=1/I$ for $i=1,...,I$ where rows represent the samples, while the column
marginals vary $min_{j}(p_{+j}^{(\alpha)})\leq p_{+j}^{(\alpha)}\leq\max
_{j}(p_{+j}^{(\alpha)})$ for $j=1,...,J$. In the next two paragraphs we
discuss two paricular cases $\alpha=1$ and $0.$

b) Benz\'{e}cri (1973b, pp. 18-27) emphasized the following fact: CA must be
applied only to \textit{homogenous} datasets; for a compositional dataset
\textbf{N }where the rows represent samples, homogeneity is obtained by
eliminating the scale of each row, changing \textbf{N} into a row stochastic
matrix RS(\textbf{N)} = ($n_{ij}/\sum_{j}n_{ij})$; this action is called
\textit{closure} within the framework of compositional data analysis. This
case corresponds to $\alpha=1.$ The underlying idea of Benz\'{e}cri's
homogenity principle is identical to Aitchison (1986)'s \textit{coherence}
principle for compositional data, because ratios of compositional parts
($n_{ij}/n_{ij_{1}})$ for $j\neq j_{1}$ or $n_{ij}/(\Pi_{j}n_{ij})^{1/J}$
eliminate the $i$th row scale. Nte that the pioneer of the Dutch school of
data science, Jan de Leeuw, referred to Multiple Correspondence Analysis (MCA)
as Homogeneity Analysis within the Gifi (1990) system.

c) We note that for sparse and extremely sparse data, the sign(\textbf{N})
corrsponds to $\alpha=0$ but it is not row stochastic; so another variant for
$\alpha=0$ is to transform sign(\textbf{N}) into a row stochastic matrix
RS(sign(N)).

\section{Applications}

We evaluate the three scale-invariant CA (and its robust L$_{1}$ variant TCA)
methods by analyzing four data sets with different sparsity indices. For
computations, we use the following two R packages: \textit{ca} by Greenacre et
al.(2022) and TaxicabCA by Allard and Choulakian (2019).

\subsection{Cups dataset}

The archaeological \textquotedblleft Cups\textquotedblright\ compositional
dataset \textbf{N} comprises 47 observations by 11 variables, originally
compiled by Baxter et al. (1990). Each of the 47 cups is described by the
percentage by weight of 11 chemical elements. Greenacre and Lewi (2009)
analyzed this dataset using log-ratio analysis (LRA), and it can be downloaded
from the R package easyCODA (Greenacre 2020). Since every entry in the Cups
dataset is strictly positive, Sinkhorn's (1964) theorem ensures that the
Sinkhorn (RAS - ipf) algorithm converges to a unique solution.

Given the positivity of all cells, Figure 1 in Greenacre and Lewi (2009)
presents the biplot from uniformly weighted LRA---also as Greenacre's
scale-invariant CA biplot---which is almost identical to the principal map
(not shown) produced by Goodman's mfCA. Below are the dispersion measures
(nearly the same size) for both principal maps.%

\begin{tabular}
[c]{|c|c|c|}\hline
& $\rho_{1}^{2}$ (variance explained) & $\rho_{2}^{2}$ (variance
explained)\\\hline
Greenacre's scale-invariant CA & $\rho_{1}^{2}=0.00833$ ($39.6\%)$ & $\rho
_{2}^{2}=0.00638\ (30.0\%)$\\\hline
Goodman's mfCA & $\rho_{1}^{2}=0.0101$ ($43.6\%)$ & $\rho_{2}^{2}=0.00673$
$(29.1\%)$\\\hline
\end{tabular}

We just note that the transformation $sign(\mathbf{N)}$ is senseless, because
$\mathbf{N}=(n_{ij})$ is nonsparse, $n_{ij}>0$.

\bigskip%
\begin{tabular}
[c]{l||ll||ll}%
\multicolumn{5}{l}{\textbf{Table 1: Sinkhorn (RAS-ipf) algorithm iteration
results.}}\\\hline
& \textit{rodent} &  & \textit{Copepods} & \\\hline
$\mathit{iteration}$ & $C2_{dist}$ & $ratio$ & $C2_{dist}$ & $ratio$\\\hline
\textit{491} & 302.96414 & 1.380883 & \ 1.98952e-13 & 1\\
\textit{492} & 257.13691 & 1.380883 & 2.819966e-13 & 1\\
\textit{493} & 302.96414 & 1.380883 & 1.98952e-13 & 1\\
\textit{494} & 257.13691 & 1.380883 & 2.819966e-13 & 1\\
\textit{495} & 302.96414 & 1.380883 & 1.98952e-13 & 1\\
\textit{496} & 257.13691 & 1.380883 & 2.819966e-13 & 1\\
\textit{497} & 302.96414 & 1.380883 & 1.98952e-13 & 1\\
\textit{498} & 257.13691 & 1.380883 & 2.819966e-13 & 1\\
\textit{499} & 302.96414 & 1.380883 & 1.98952e-13 & 1\\
\textit{500} & 257.13691 & 1.380883 & 2.819966e-13 & 1\\\hline
\end{tabular}

\subsection{\textit{Rodent} ecological abundance dataset}

We analyze a dataset of rodent abundances ($\mathbf{N}$) comprising 9 species
observed in 28 California cities. Originally published by Bolger et al. (1997)
and CA analyzed by Quinn and Keough (2002, p. 460, Figure 17.5). This dataset
is available in the TaxicabCA R package (Allard \& Choulakian, 2019).

The apparent sparsity of $(\mathbf{N)=\ }66.27\%$\textbf{;} the size of
$\mathbf{N}_{merged}$ is $21\times9$ and the CA\textbf{ }sparsity\textbf{ of
}$(\mathbf{N)=}$ $58.7\%$; $S(21,9)=21+9-3=27$, so the adjusted sparsity of
$(\mathbf{N)=\ }58.7/(1-27/(21\ast9))=68.48\%$.

Choulakian (2017) analyzed it by comparing the CA and TCA maps, and showed
that it has quasi 2-blocks diagonal structure (or partly decomposable) as
shown in Table 3 by Benz\'{e}cri's empirical criterion stated in Subsection
2.4, see ca(\textbf{N})\$sv, ca(RS(\textbf{N}))\$sv, ca(sign(\textbf{N}))\$sv
and ca(RS(sign(\textbf{N})))\$sv, where RS means row stochastic:%

\begin{tabular}
[c]{|c|c|c|c|c|c|c|}\hline
& $\rho_{1}$ & $\rho_{2}$ & $\rho_{3}$ & $\rho_{4}$ & $\rho_{5}$ & $\rho_{6}%
$\\\hline
ca(\textbf{N})\$sv & 0.8639 & 0.6776 & 0.5362 & 0.3909 & 0.1889 &
0.1568\\\hline
ca(RS(N))\$sv & 0.9554 & 0.8122 & 0.6211 & 0.5251 & 0.2009 & 0.1629\\\hline
ca(sign(\textbf{N}))\$sv & 0.8167 & 0.5990 & 0.4458 & 0.4106 & 0.2605 &
0.2171\\\hline
ca(RS(sign(\textbf{N})))\$sv & 0.8885 & 0.7704 & 0.5299 & 0.4795 & 0.3759 &
0.2622\\\hline
\end{tabular}

\subsubsection{CA of $sign\mathbf{(N)}$}

The close correspondence of dispersion values of ca(\textbf{N})\$sv and
ca(sign(\textbf{N}))\$sv indicates that the principal maps (not shown) from CA
on raw abundances and on their sign transformation are highly similar.

\subsubsection{mfCA}

Choulakian and Mahdi (2024) applied Goodman's mfCA and Taxicab mfCA (mfTCA)
after using the ipfr R package (Ward \& Macfarlane, 2020), which regularizes
zeros by replacing them with a tiny positive constant to guarantee unique
convergence of Sinkhorn's algorithm. In our approach, zeros remain unchanged,
causing the Sinkhorn (RAS-ipf) algorithm to oscillate between two bistochastic
solutions, see Table 1, when run for a fixed 500 iterations due to the
underlying quasi 2-blocks diagonal structure (or partly decomposable) of
\textbf{N}. Both solutions produce the same four blocks but differ in their
uniform row and column marginals, see Table 2. By Benz\'{e}cri's Theorem 1, we
validate convergence to the same four blocks by comparing CA singular values
for the scaled matrix $\mathbf{D}$ at iterations 499 and 500:%

\begin{tabular}
[c]{|c|c|c|c|c|c|c|c|c|}\hline
& $\rho_{1}$ & $\rho_{2}$ & $\rho_{3}$ & $\rho_{4}$ & $\rho_{5}$ & $\rho_{6}$
& $\rho_{7}$ & $\rho_{8}$\\\hline
iter = 500: ca(\textbf{D})\$sv & 1 & 1 & 1 & 0.8052 & 0.7174 & 0.6336 &
0.4936 & 0.2558\\\hline
iter = 499: ca(\textbf{D})\$sv & 1 & 1 & 1 & 0.8052 & 0.7174 & 0.6336 &
0.4936 & 0.2558\\\hline
\end{tabular}

Identical dispersion profiles at successive iterations provide a practical
convergence check for large and sparse datasets.

The final scaled matrix \textbf{D} decomposes into four blocks (Case 2), as
displayed in Table 3. We proceed to interpret these blocks in terms of
species--city associations.%

\begin{tabular}
[c]{lllll}%
\multicolumn{5}{l}{\textbf{Table 2: Row and column uniform marginals of 4
blocks of\ \textit{rodent} data.}}\\\hline\hline
\textit{iteration 500 (499) } &  & \multicolumn{3}{l}{$C2_{dist}%
=257.14\ (302.96)$}\\\hline
\textit{100*colPROBA}$\mathit{(c}_{\beta})$ &
\multicolumn{1}{|l}{36.23(\textit{4.76})} & 29.96(\textit{6.71}) &
\multicolumn{1}{|l}{5.25(\textit{13.13})} & 3.76(\textit{11.45})\\
\textit{columns } & \multicolumn{1}{|l}{\textbf{1}} & \textbf{2} &
\multicolumn{1}{|l}{\textbf{3,4,5,6,8}} & \textbf{7,9}\\\hline
\textit{100*rowPROBA}$\mathit{(r}_{\beta})$ &
\multicolumn{1}{|l}{6.04\textit{(0.79})} & 4.28(\textit{0.96}) &
2.51(\textit{7.63}) & 2.19(\textit{5.47})\\
rows & \multicolumn{1}{|l}{\textbf{24,21,17,}} & \textbf{25,22,16,15} &
\textbf{28,27,26,23} & \textbf{6,4,2}\\
& \multicolumn{1}{|l}{\textbf{14,10,9}} & \textbf{11,8,7} &
\textbf{20,19,18,13} & \\
& \multicolumn{1}{|l}{} &  & \textbf{12,5,3,1} & \\\hline
\end{tabular}

\paragraph{Interpretation of Table 3}

In Table 3, the array $\mathbf{T}=(t_{ij}=d_{ij\alpha}|\mathbf{n}_{ij}),$
represents both matrices the original \textit{rodent} contingency table
$\mathbf{N}=(n_{ij})$ and the 4 bistochastic blocks $\mathbf{T}_{\alpha
}=(d_{ij\alpha})$ for $\alpha=1,2,3,4$, where each bistochastic block is
aligned on both left and right sides by double lines. For example,%
\begin{align*}
\mathbf{T}_{1}  &  =(t_{ij}=d_{ij1}|\mathbf{n}_{ij})\\
&  =\
\begin{tabular}
[c]{||l||}%
1%
$\vert$%
\textbf{1}\\
1%
$\vert$%
\textbf{3}\\
1%
$\vert$%
\textbf{2}\\
1%
$\vert$%
\textbf{1}\\
1%
$\vert$%
\textbf{3}\\
1%
$\vert$%
\textbf{1}%
\end{tabular}
\end{align*}
which is of size $6\times1;$ \textbf{T}$_{2}$ is of size $7\times1;$
\textbf{T}$_{3}$ is of size $12\times5$ and \textbf{T}$_{4}$ is of size
$3\times2$. A blank cell represents (0%
$\vert$%
\textbf{0), }such as\textbf{ }t$_{12}=$(0%
$\vert$%
\textbf{0).}

The interpretation of $\mathbf{T}_{1}$ is: rodent 1 locally characterizes the
sites (24, 1, 21, 10, 9, 14). Note that there is an observed count of 3 rodent
1s in the site 20, but rodent 1 does not locally characterize site 20; site 20
belongs to the third bistochastic block which is locally associated with the
rodents (6, 8, 3, 5, 4). Similar interpretations also apply to the other three
bistochastic matrices. So, the 4 bistochastic blocks can be considered as a
classification of the abundance data set \textbf{N}.

We also note the following: number of zero valued cells is 167 in \textbf{N
}(or in sign(\textbf{N})); while the number of zeros is 196 in the scaled 4
blocks diagonal bistochastic submatrices in Table 3.\bigskip%

\begin{tabular}
[c]{llllllllll}%
\multicolumn{10}{l}{\textbf{Table 3: 4 blocks diagonal bistochastic
\textit{rodent} abundance data.}}\\\hline
$sites$ & \textit{rod1} & \textit{rod2} & \textit{rod6} & \textit{rod8} &
\textit{rod3} & \textit{rod5} & \textit{rod4} & \textit{rod7} & \textit{rod9}%
\\\hline
\textit{24} & \multicolumn{1}{||l}{1%
$\vert$%
\textbf{1}} & \multicolumn{1}{||l}{} &  &  &  &  &  &  & \\
\textit{17} & \multicolumn{1}{||l}{1%
$\vert$%
\textbf{3}} & \multicolumn{1}{||l}{} &  &  &  &  &  &  & \\
\textit{21} & \multicolumn{1}{||l}{1%
$\vert$%
\textbf{2}} & \multicolumn{1}{||l}{0%
$\vert$%
\textbf{1}} &  &  &  &  &  &  & \\
\textit{10} & \multicolumn{1}{||l}{1%
$\vert$%
\textbf{1}} & \multicolumn{1}{||l}{0%
$\vert$%
\textbf{2}} &  &  &  &  &  &  & \\
\textit{9} & \multicolumn{1}{||l}{1%
$\vert$%
\textbf{3}} & \multicolumn{1}{||l}{0%
$\vert$%
\textbf{8}} &  &  &  &  &  &  & \\
\textit{14} & \multicolumn{1}{||l}{1(\textbf{1)}} & \multicolumn{1}{||l}{0%
$\vert$%
\textbf{3}} &  &  &  &  &  &  & \\
\textit{7} & \multicolumn{1}{|l}{} & \multicolumn{1}{||l}{1%
$\vert$%
\textbf{11}} & \multicolumn{1}{||l}{} &  &  &  &  &  & \\
\textit{11} & \multicolumn{1}{|l}{} & \multicolumn{1}{||l}{1%
$\vert$%
\textbf{9}} & \multicolumn{1}{||l}{} &  &  &  &  &  & \\
\textit{15} & \multicolumn{1}{|l}{} & \multicolumn{1}{||l}{1%
$\vert$%
\textbf{11}} & \multicolumn{1}{||l}{} &  &  &  &  &  & \\
\textit{25} & \multicolumn{1}{|l}{} & \multicolumn{1}{||l}{1%
$\vert$%
\textbf{5}} & \multicolumn{1}{||l}{} &  &  &  &  &  & \\
\textit{22} & \multicolumn{1}{|l}{} & \multicolumn{1}{||l}{1%
$\vert$%
\textbf{3}} & \multicolumn{1}{||l}{} &  &  &  &  &  & \\
\textit{8} & \multicolumn{1}{|l}{} & \multicolumn{1}{||l}{1%
$\vert$%
\textbf{16}} & \multicolumn{1}{||l}{} &  &  &  &  &  & \\
\textit{16} & \multicolumn{1}{|l}{} & \multicolumn{1}{||l}{1%
$\vert$%
\textbf{4}} & \multicolumn{1}{||l}{} &  &  &  &  &  & \\
\textit{1} & \multicolumn{1}{|l}{} & 0%
$\vert$%
\textbf{13} & \multicolumn{1}{||l}{0.97%
$\vert$%
\textbf{2}} &  & 0.19%
$\vert$%
\textbf{3} & 1.41%
$\vert$%
\textbf{1} & 2.43%
$\vert$%
\textbf{1} & \multicolumn{1}{||l}{} & \\
\textit{20} & \multicolumn{1}{|l}{0%
$\vert$%
\textbf{3}} &  & \multicolumn{1}{||l}{} &  & 2.09%
$\vert$%
\textbf{27} &  & 2.91%
$\vert$%
\textbf{1} & \multicolumn{1}{||l}{} & \\
\textit{12} & \multicolumn{1}{|l}{} & 0%
$\vert$%
\textbf{3} & \multicolumn{1}{||l}{0.85%
$\vert$%
\textbf{16}} & 3.38%
$\vert$%
\textbf{7} & 0.01%
$\vert$%
\textbf{1} & 0.76%
$\vert$%
\textbf{5} &  & \multicolumn{1}{||l}{} & \\
\textit{3} & \multicolumn{1}{|l}{} & 0%
$\vert$%
\textbf{4} & \multicolumn{1}{||l}{2.30%
$\vert$%
\textbf{9}} &  & 1.22%
$\vert$%
\textbf{36} & 1.48%
$\vert$%
\textbf{2} &  & \multicolumn{1}{||l}{} & \\
\textit{13} & \multicolumn{1}{|l}{} & 0%
$\vert$%
\textbf{4} & \multicolumn{1}{||l}{2.09%
$\vert$%
\textbf{12}} &  & 0.90%
$\vert$%
\textbf{39} & 2.01%
$\vert$%
\textbf{4} &  & \multicolumn{1}{||l}{} & \\
\textit{28} & \multicolumn{1}{|l}{} &  & \multicolumn{1}{||l}{0.68%
$\vert$%
\textbf{1}} &  & 0.91%
$\vert$%
\textbf{10} &  & 3.41%
$\vert$%
\textbf{1} & \multicolumn{1}{||l}{} & \\
\textit{18} & \multicolumn{1}{|l}{} & 0%
$\vert$%
\textbf{2} & \multicolumn{1}{||l}{0.78%
$\vert$%
\textbf{14}} & 2.03%
$\vert$%
\textbf{4} & 0.58%
$\vert$%
\textbf{78} & 1.61%
$\vert$%
\textbf{10} &  & \multicolumn{1}{||l}{} & \\
\textit{5} & \multicolumn{1}{|l}{} & 0%
$\vert$%
\textbf{2} & \multicolumn{1}{||l}{0.50%
$\vert$%
\textbf{16}} &  & 0.26%
$\vert$%
\textbf{63} & 0.99%
$\vert$%
\textbf{11} & 3.26%
$\vert$%
\textbf{21} & \multicolumn{1}{||l}{} & \\
\textit{23} & \multicolumn{1}{|l}{} &  & \multicolumn{1}{||l}{1.25%
$\vert$%
\textbf{8}} & 2.85%
$\vert$%
\textbf{2} &  & 0.90%
$\vert$%
\textbf{2} &  & \multicolumn{1}{||l}{} & \\
\textit{26} & \multicolumn{1}{|l}{} &  &
\multicolumn{1}{||l}{1.71(\textbf{11)}} & 2.84(\textbf{2)} & 0.45(\textbf{22)}
&  &  & \multicolumn{1}{||l}{} & \\
\textit{27} & \multicolumn{1}{|l}{} &  & \multicolumn{1}{||l}{0.88%
$\vert$%
\textbf{9}} & 0.90%
$\vert$%
\textbf{1} & 0.38%
$\vert$%
\textbf{29} & 2.84%
$\vert$%
\textbf{10} &  & \multicolumn{1}{||l}{} & \\
\textit{19} & \multicolumn{1}{|l}{} &  & \multicolumn{1}{||l}{} &  & 5%
$\vert$%
\textbf{1} &  &  & \multicolumn{1}{||l}{} & \\
\textit{6} & \multicolumn{1}{|l}{} & 0%
$\vert$%
\textbf{1} & 0%
$\vert$%
\textbf{8} & 0%
$\vert$%
\textbf{2} & 0%
$\vert$%
\textbf{48} & 0%
$\vert$%
\textbf{12} & 0%
$\vert$%
\textbf{35} & \multicolumn{1}{||l}{1.65%
$\vert$%
\textbf{12}} & \multicolumn{1}{l||}{0.35%
$\vert$%
\textbf{2}}\\
\textit{2} & \multicolumn{1}{|l}{} & 0%
$\vert$%
\textbf{1} & 0%
$\vert$%
\textbf{16} & 0%
$\vert$%
\textbf{2} & 0%
$\vert$%
\textbf{57} & 0%
$\vert$%
\textbf{9} & 0%
$\vert$%
\textbf{65} & \multicolumn{1}{||l}{1.35%
$\vert$%
\textbf{8}} & \multicolumn{1}{l||}{0.65%
$\vert$%
\textbf{3}}\\
\textit{4} & \multicolumn{1}{|l}{} & 0%
$\vert$%
\textbf{4} & 0%
$\vert$%
\textbf{30} & 0%
$\vert$%
\textbf{18} & 0%
$\vert$%
\textbf{53} & 0%
$\vert$%
\textbf{5} & 0%
$\vert$%
\textbf{1} & \multicolumn{1}{||l}{} & \multicolumn{1}{l||}{2%
$\vert$%
\textbf{3}}\\\hline
\end{tabular}

\bigskip

\subsection{Copepods dataset}

We analyze the \textit{Copepods} (sea insects)\textit{ }dataset \textbf{N }of
size $42\times40$ found in Greenacre's \textbf{CODAinPractice} website, which
is analyzed by Graeve and Greenacre (2020). It comprises $I=42$ specimens,
each described by $J=40$ fatty acids. Every specimen was captured in one of
three seasons:

- Summer (s): 22 specimens

- Spring (sp): 12 specimens

- Winter (w): 8 specimens

The study's objective is to determine whether these fatty-acid profiles can
distinguish among the three seasons.

\subsubsection{Sparsity of the data}

The dataset has an apparent sparsity (and CA sparsity) of 11.13\%. Adjusted
sparsity of 11.69\% by using the formula $\mathbf{\ }%
11.13/(1-(42+40-2)/(42\ast40)).$

\subsubsection{mfCA and mfTCA results}

We applied the Sinkhorn (RAS-ipf) algorithm to scale the original data matrix
\textbf{N} into a unique bistochastic matrix \textbf{D}, see Table 1. After
500 iterations, the resulting probability matrix $\mathbf{Q}=(q_{ij}%
=\frac{d_{ij}}{IJ}=\frac{d_{ij}}{589})$ has uniform row and column sums:
$q_{i+}=\frac{1}{I}=0.02380952$ and $q_{+j}=\frac{1}{J}=0.025.$

Figures 1 and 2 display the maps from Goodman's mfCA and mfTCA of $\mathbf{Q}%
$, respectively. The mfCA map Figure 1  is challenging to interpret because CA
is sensitive to outliers. In contrast, Figure 2 reveals a clear separation of:

- Summer (s) points lie on the right side of the plot

- Spring (sp) and Winter (w) points on the left side, with no clear separation
between (sp) and (w) points. This is due to the fact that the bistochastic
matrix $\mathbf{D}$ has changed the \textit{internal order structure} of the
original row stochastic matrix $\mathbf{N}$.

(Insert Figures 1 and 2)

\begin{figure}
  \includegraphics{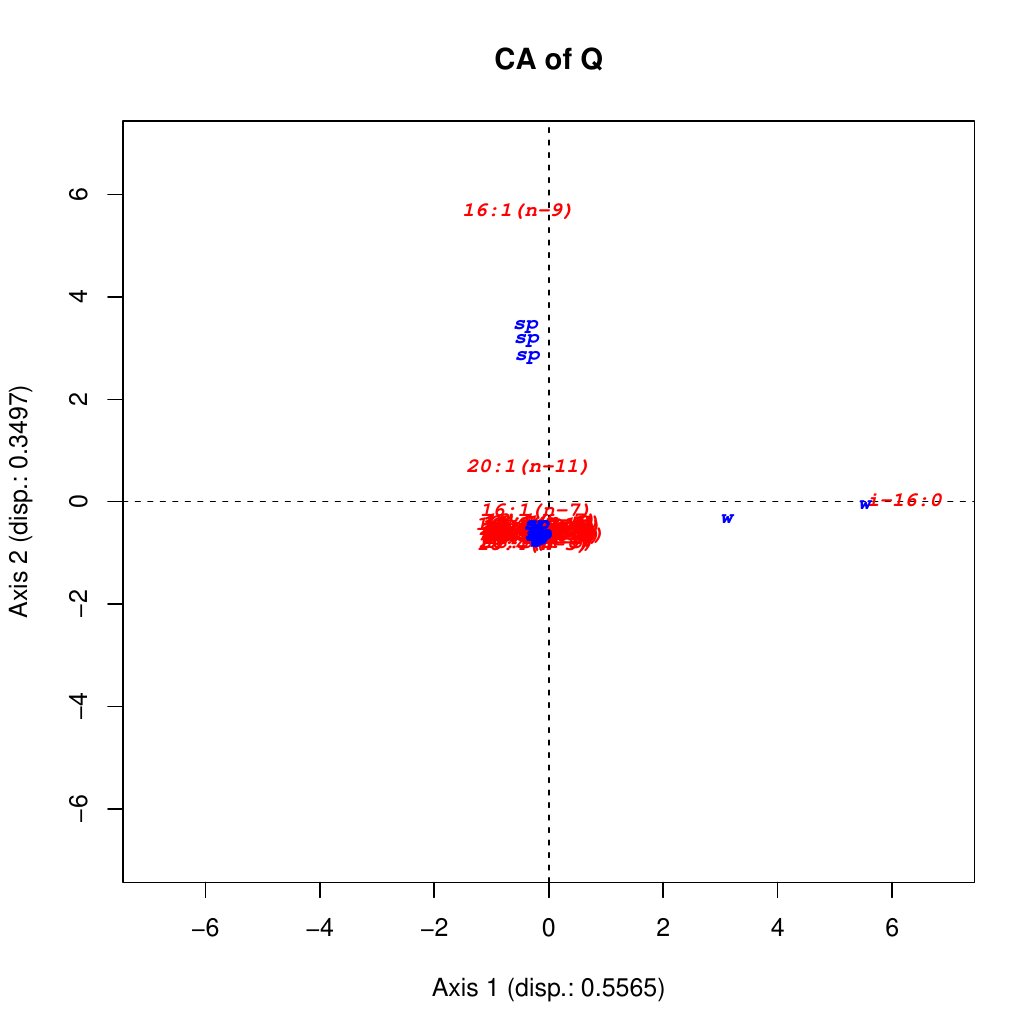}
\caption{mfCA of Copepods dataset.}
\label{fig:1}       
\end{figure}
%
%

\begin{figure}
  \includegraphics{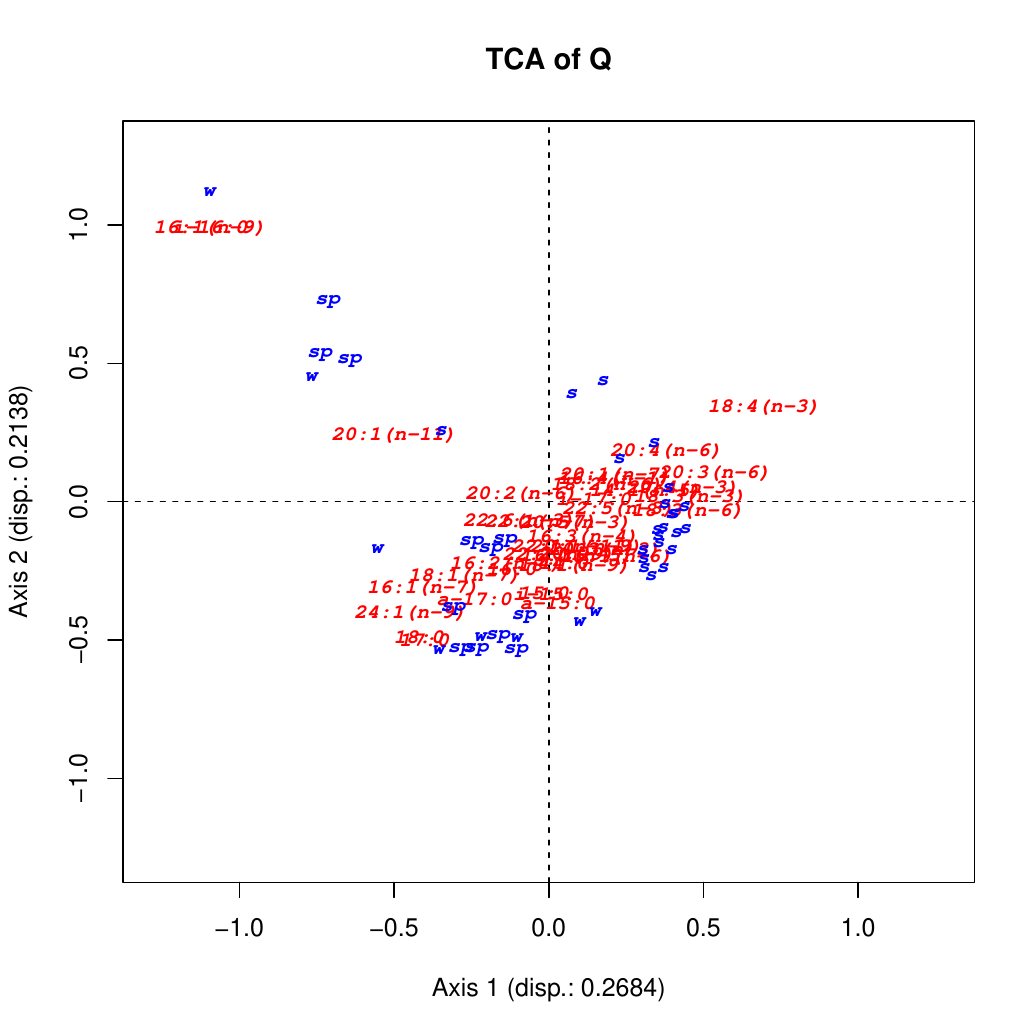}
\caption{mfTCA of Copepods dataset.}
\label{fig:2}       
\end{figure}
%
%

\subsubsection{CA and TCA of the row stochastic dataset $\mathbf{N}$}

Benz\'{e}cri (1973b, pp. 18-27) emphasized the following fact: CA must be
applied only to \textit{homogenous} datasets; for compositional datasets this
criterion was shown to be successful by Baxter et al. (1990). The available
compositional dataset $\mathbf{N}=(n_{ij})$ being row stochastic is
homogenous, that is, $\sum_{j}n_{ij}=100$ for $i=1,...,42$, while the column
sums vary \ $0.2543986\leq\sum_{i}n_{ij}\leq558.1079$. Figures 3 and 4 display
the CA and TCA maps of $\mathbf{N}$ respectively, where we clearly see the
separation of the three seasonal specimens.

(Insert Figures 3 and 4)

\begin{figure}
  \includegraphics{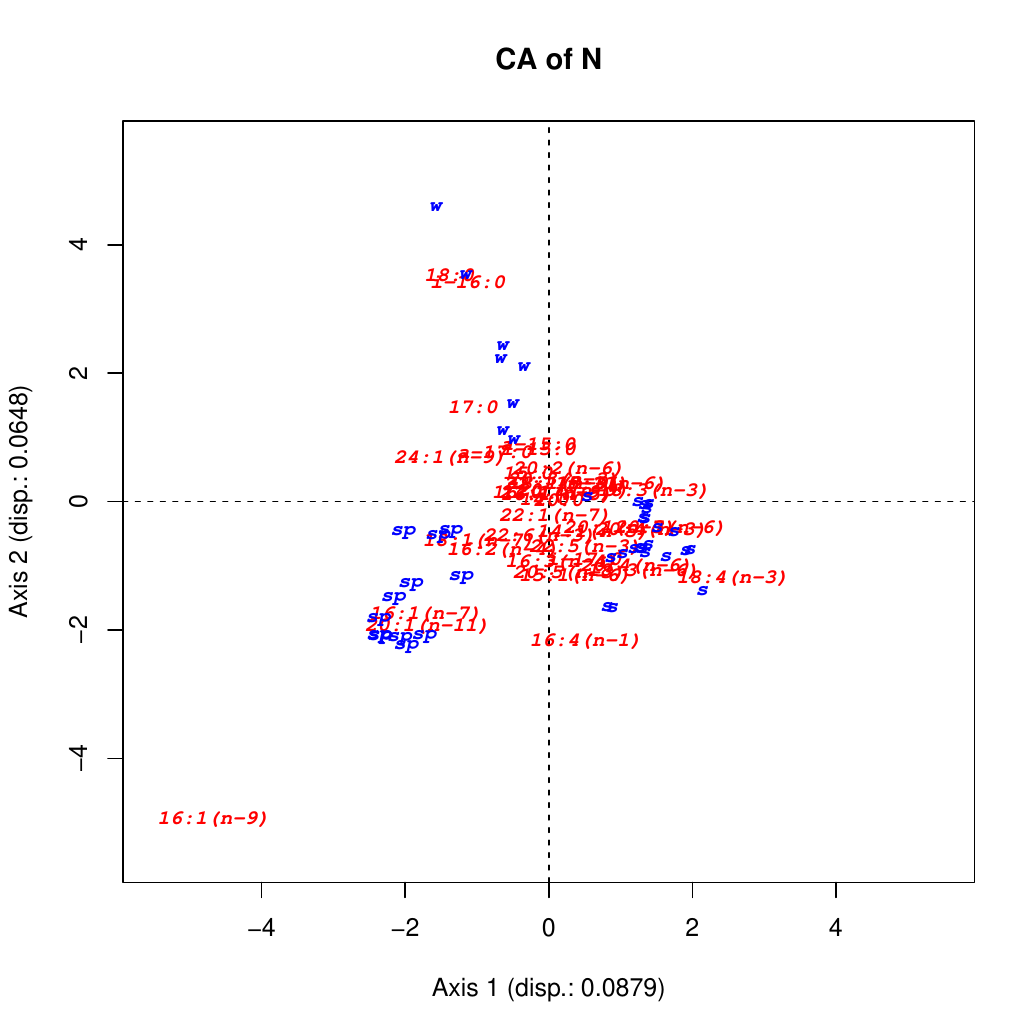}
\caption{CA of RowStocastic Copepods dataset.}
\label{fig:3}       
\end{figure}
%
%

\begin{figure}
  \includegraphics{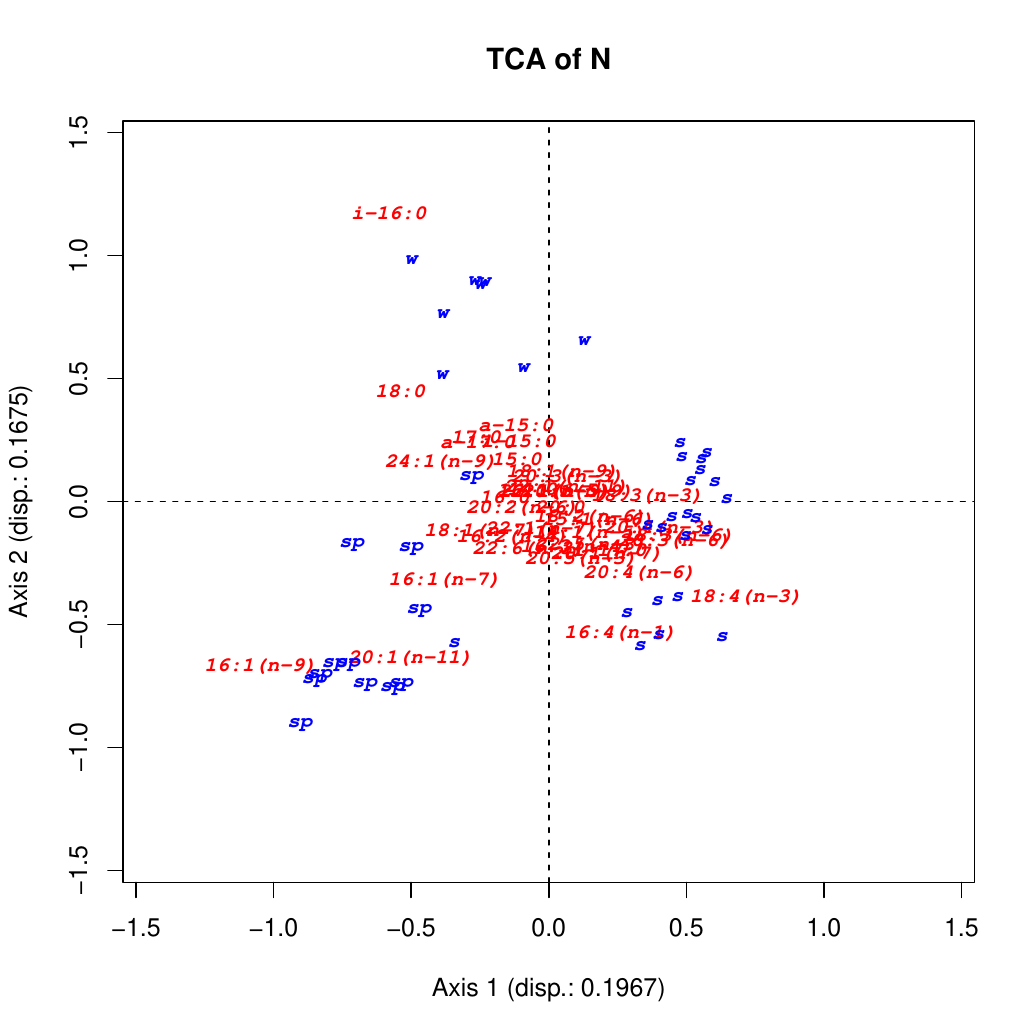}
\caption{TCA of RowStocastic Copepods dataset.}
\label{fig:4}       
\end{figure}
%
%

\subsection{Sacred Books dataset}

Sah and Fokou\'{e} (2019) constructed an extremely sparse contingency table
$\mathbf{N}$ of size $I\times J=590\times8266$. Here, 590 denotes fragments of
chapters extracted from English translations of eight sacred books, and 8 266
is the number of distinct words (tokens) across the chosen fragments.

The dataset includes fragments from four Biblical books and four
extreme-Orient texts. The symbols in parentheses will label each fragment on
the CA and TCA maps:%

\begin{tabular}
[c]{c|c|c}%
\textbf{Text} & \textbf{Symbol} & \textbf{Number of chapters}\\\hline
Four Noble Truths of Buddhism & B & 45\\
Tao Te Ching & T & 81\\
Upanishads & U & 162\\
Yogasutras & Y & 189\\
Book of Proverbs & P & 31\\
Book of Ecclesiastes & E & 12\\
Book of Ecclesiasticus & e & 50\\
Book of Wisdom & W & 19\\\hline
\end{tabular}

\subsubsection{Sparsity and dataset characteristics}

- The total number of Oriental chapters is 478, with 30 252 words counted. The
total number of Biblical chapters is 112, with 30 355 words counted. So, on
average, the Oriental fragments are much sparser than the Biblical ones.

- The apparent sparsity of \textbf{N} is 99.14\%. After merging and
eliminating the one empty fragment (chapter 14 of the B text), we obtain the
CA sparsity of (\textbf{N}) = apparent sparsity $(\mathbf{N}_{merged}%
)=98.65\%,$ where $\mathbf{N}_{merged}$ is of size $589\times4864$. Note that
the marginal count of the words in the chosen fragment of chapter 14 of the B
book is null; so in CA and TCA this row is eliminated. This extreme sparsity
(often described as high dimensionality) poses challenges for many analysis methods.

\subsubsection{Dimension-reduction methods reviewed by Ma, Sun \& Zou (2023)}

Ma, Sun and Zou visualized the dataset with twelve classic methods and their
newly proposed meta-visualization method:

1) Principal component analysis (PCA) (Hotelling 1933)

2) Multi-dimensional scaling (MDS) (Torgerson 1952)

3) Non-metric MDS (iMDS) (Kruskal 1964)

4) Sammon's mapping (Sammon) (Sammon 1969)

5) Kernel PCA (kPCA) (Scholkopf, Smola and Muller 1998)

6) Locally linear embedding (LLE) (Roweis and Saul 2000)

7) Isomap (Tenenbaum, Silva and Langford 2000)

8) Hessian LLE (HLLE) (Donoho and Grimes 2003)

9) Laplacian eigenmap (LEIM) (Belkin and Niyogi 2003)

10) t-SNE (van der Maaten and Hinton 2008)

11) Uniform manifold approximation and projection for dimension reduction
(UMAP) (McInnes, Healy and Melville 2018)

12) Potential of Heat-diffusion for Affinity-based Trajectory Embedding
(PHATE) (Moon et al. 2019)

13) Meta-visualization (meta-spec) (Ma, Sun and Zou 2022, 2023)

\subsubsection{CA and TCA of the count compositional dataset N}

When analyzing the dataset \textbf{N}, two interpretations arise:

- If \textbf{N} represents the full population of all sacred-text
chapter-fragments, CA's use of actual marginal weights accurately reflects
sacred-texts structure.

- If \textbf{N} is viewed as a count compositional sample (akin to the Cups,
rodent and Copepods datasets), scale-invariant CA methods are more
appropriate. Given that the contingency table is extremely sparse, we can have
three variants:

a) Figures 5 and 6 display the CA and TCA maps of sign(\textbf{N}) for the 589
fragments. The CA map is difficult to interpret, while the TCA map closely
resembles the PHATE and Meta-Spec visualizations, clearly separating Biblical
fragments (left) from Oriental fragments (right).

(Insert Figures 5 \& 6 here)

%
\begin{figure*}
  \includegraphics[width=1.35\textwidth]{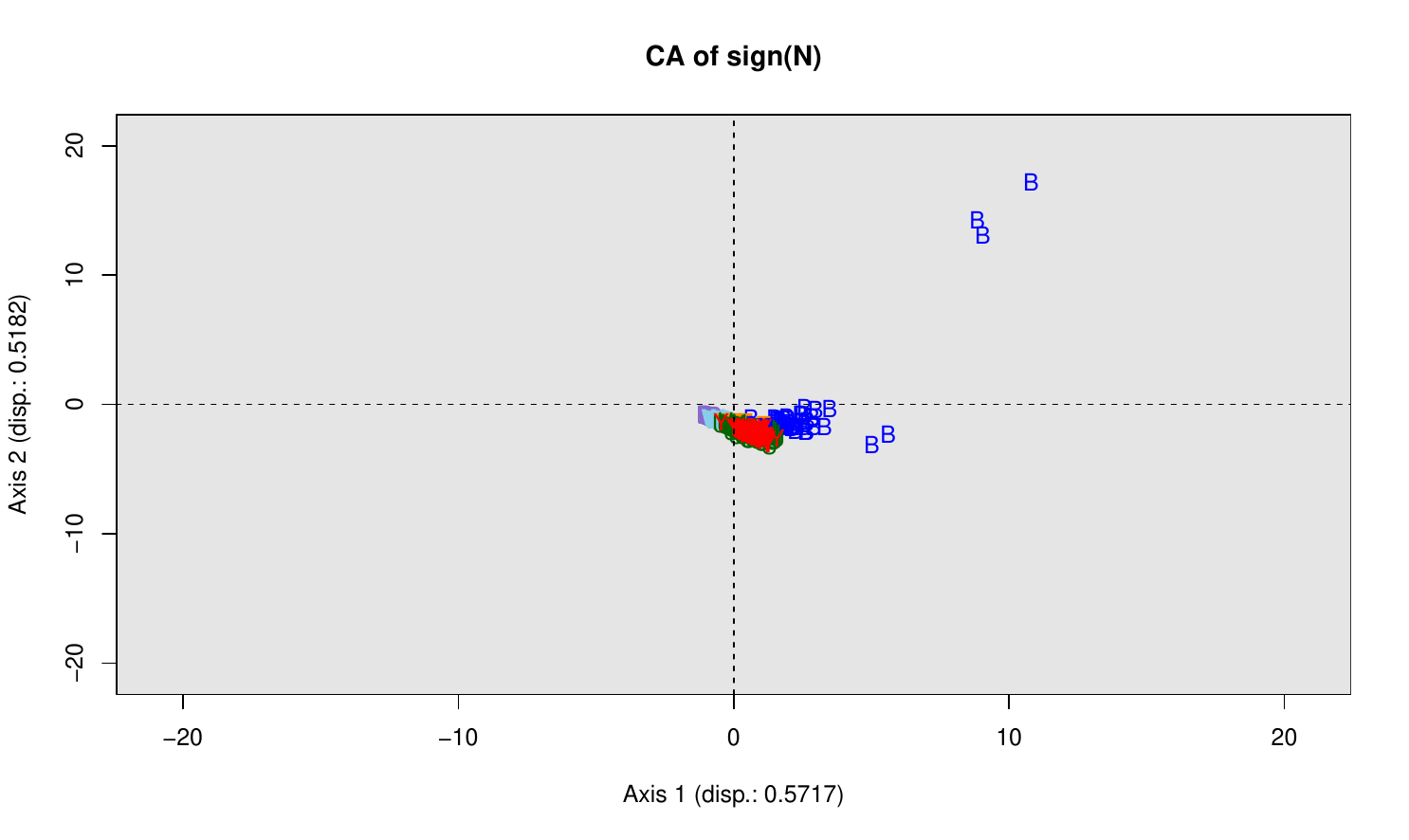}
\caption{CA of sign(SacredBooks) dataset.}
\label{fig:5}       
\end{figure*}

\begin{figure}
  \includegraphics{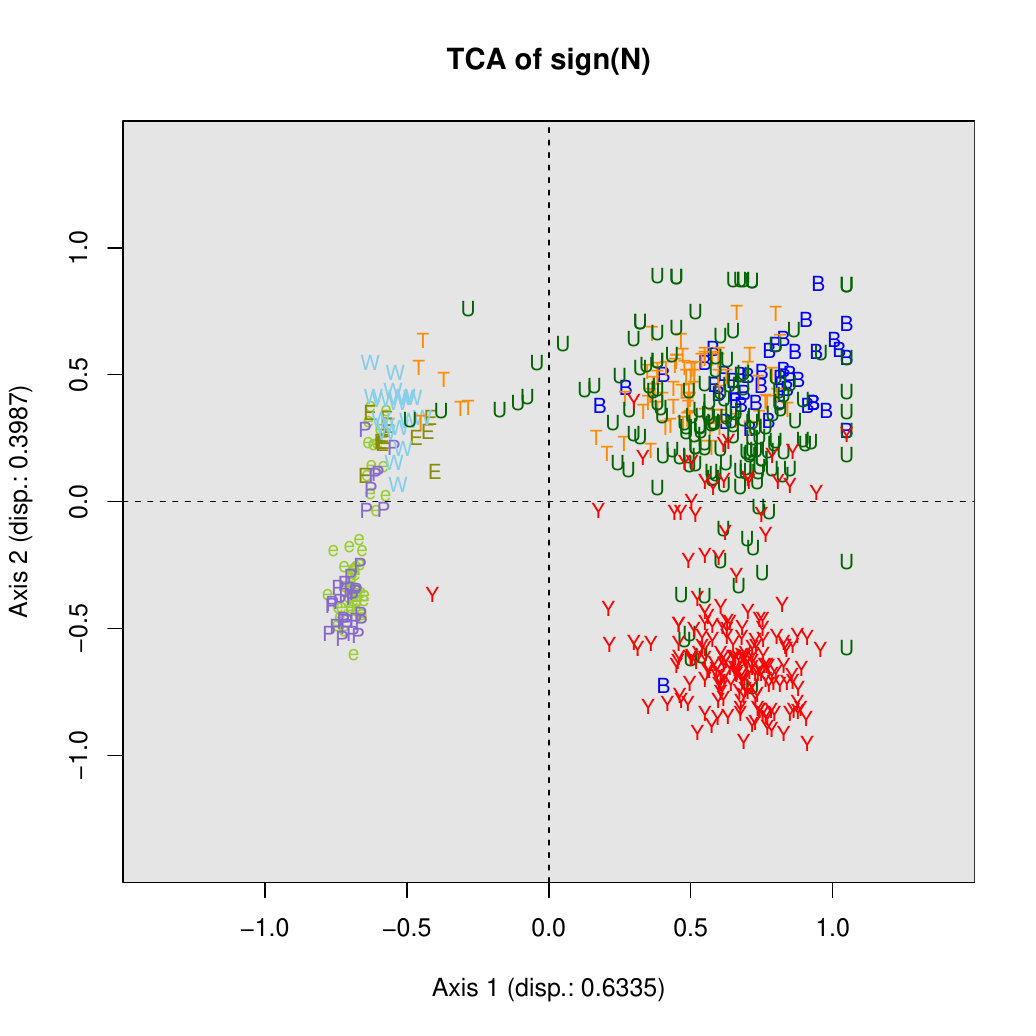}
\caption{TCA of sign(SacredBooks) dataset.}
\label{fig:6}       
\end{figure}
%
%

b) Figure 7 represents the TCA map of Benz\'{e}cri's homogenous row stochastic
transformed dataset RS(\textbf{N)}, where we clearly note the dominance of the
Oriental fragments (distributed in three quadrants) over the Biblical
fragments (concentrated in one quarter) due to the fact that number of
Oriental chapters is 478, while the number of Biblical chapters is 112.

(Insert Figure 7 here)

\begin{figure}
  \includegraphics{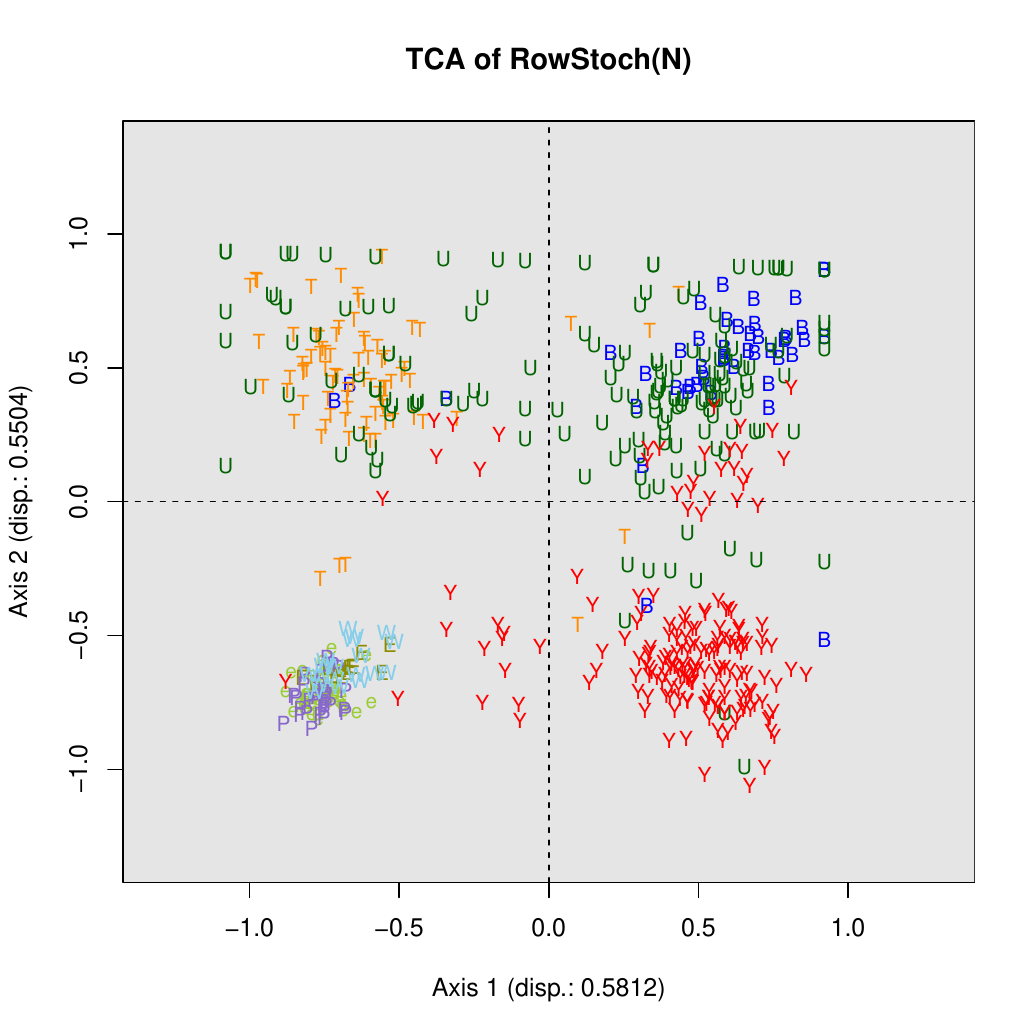}
\caption{TCA of RowStochastic (Sacred Books) dataset.
}
\label{fig:7}       
\end{figure}
%
%

c) Figure 8 represents row stochastic transformed dataset RS(sign(N)), where
we clearly note its close similarity with Figure 7, with the only difference
is the change of positions of the 44 B points.

(Insert Figure 8 here)

\begin{figure}
  \includegraphics{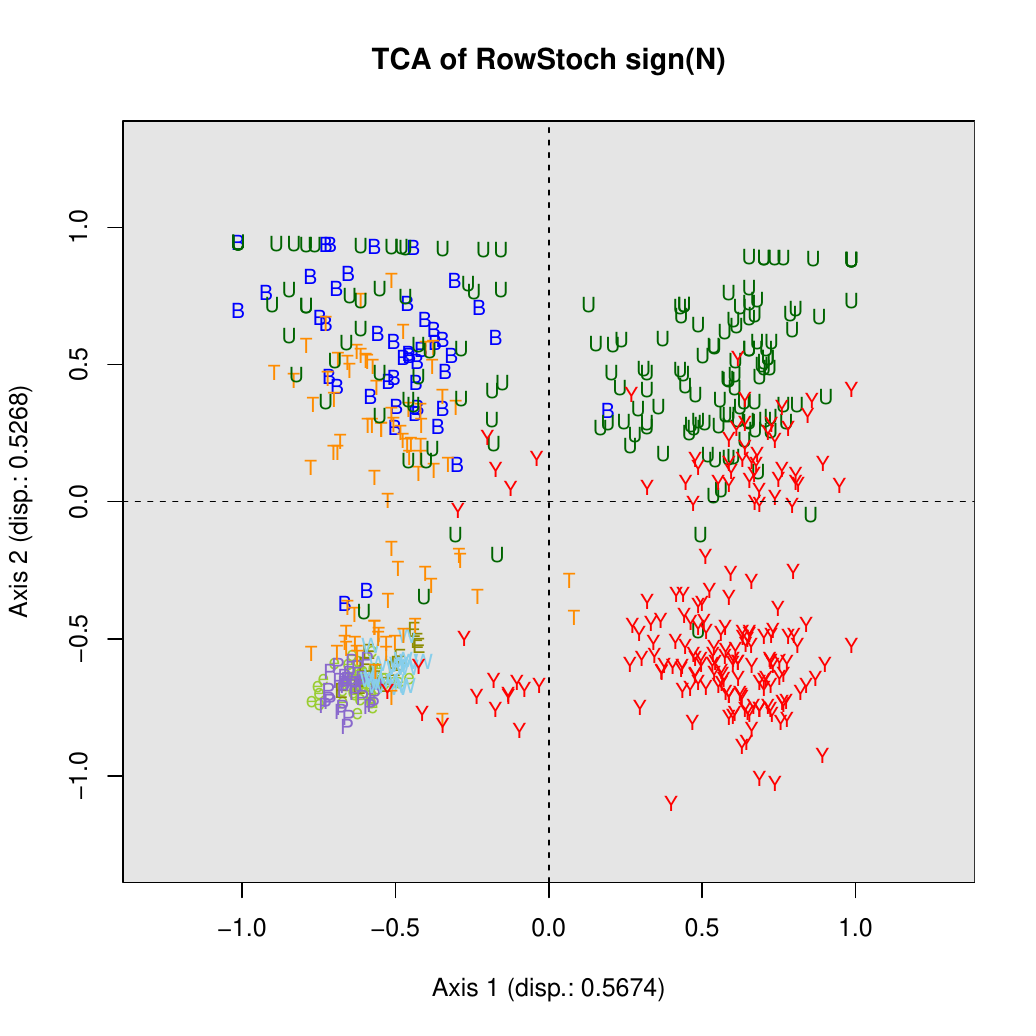}
\caption{TCA of RowStochastic sign(Sacred Books) dataset.
}
\label{fig:8}       
\end{figure}

\subsubsection{mfCA limitations}

Table 4 presents the final ten iterations of the Sinkhorn (RAS-ipf) algorithm
applied to \textbf{N}. Although the algorithm converges to two
solutions---similar to the rodent dataset in Table 1---the iteration at step
1499 in Table 4\ yields $C2_{dist}=1793232$, indicating at least 111 diagonal
blocks in the underlying structure, see Appendix. This enormous value
demonstrates that both mfCA and mfTCA struggle with extremely sparse datasets
and are therefore not recommended here.%

\begin{tabular}
[c]{l||ll}%
\multicolumn{3}{l||}{\textbf{Table 4: Sinkhorn (RAS-ipf) algorithm iteration
results.}}\\\hline
& \textit{Sacred Books} & \\\hline
$\mathit{iteration}$ & $C2_{dist}$ & $ratio$\\\hline
1\textit{491} & 1.793233e+06 & 1.002523\\
1\textit{492} & 1.855228e+06 & 1.002523\\
1\textit{493} & 1.793233e+06 & 1.002523\\
1\textit{494} & 1.855228e+06 & 1.002523\\
1\textit{495} & 1.793233e+06 & 1.002523\\
1\textit{496} & 1.855228e+06 & 1.002523\\
1\textit{497} & 1.793232e+06 & 1.00253\\
1\textit{498} & 1.855227e+06 & 1.002523\\
1\textit{499} & 1.793232e+06 & 1.002523\\
1\textit{500} & 1.855227e+06 & 1.002523\\\hline
\end{tabular}

\section{Conclusion}

We conclude our exploration by summarizing three key takeaways:

First, we analyzed four datasets of varying sparsity using three
scale-invariant CA methods and one row stochatic CA method. For each dataset,
we selected the combination of methods best suited to its sparsity level.%

\begin{tabular}
[c]{c|c|c}%
\textbf{Dataset} & \textbf{Sparsity \%} & \textbf{Methods applied}\\\hline
Cups (nonsparse) & 0 & Greenacre \& Goodman\\
Copepods (moderately sparse) & 11.69 (adjusted) & Goodman \& RS\\
Rodents (sparse) & 68.48 (adjusted) & sign-transform \& RS\\
Sacred Books (extremely sparse) & 98.65\% (CA-sparsity) & sign-transform \&
RS\\\hline
\end{tabular}

Second, applying the Sinkhorn (RAS-ipf) algorithm to a nonnegative dataset may
alter its internal order structure, thereby emphasizing the local micro-level
structure in Goodman's mfCA approach.

Third, by applying both CA and TCA side by side, we:

- Capture complementary dimensions of association in contingency tables and
compositional data

- Enhance robustness to sparsity and extreme values

- Provide richer visualization and interpretation for exploratory analysis

Overall, combining the three scale-invariant CA/TCA variants and one row
stochatic CATCA method offers a versatile toolkit for uncovering structure
across compositional datasets from nonsparse to extremely sparse.\bigskip

\bigskip

$\ \bigskip$\textbf{Declarations}

\textit{Funding:} Partial funding to Choulakian is provided by the Natural
Sciences and Engineering Research of Canada (Grant no. RGPIN-2017-05092).

\textit{Data availability:} The four real datasets used in this paper are
available online.

\bigskip

\textbf{References}

Aitchison J (1986) \textit{The Statistical Analysis of Compositional Data}.
London: Chapman and Hall

Allard J, Choulakian V (2019) \textit{Package TaxicabCA in R}. Available at:

https ://CRAN.R-project.org/package=TaxicabCA

Baxter MJ, Cool HM and Heyworth MP (1990) Principal component and
correspondence analysis of compositional data: some similarities.
\textit{Journal of Applied Statistics} 17, 229--235

Beh E, Lombardo R (2014) \textit{Correspondence Analysis: Theory, Practice and
New Strategies}. N.Y: Wiley

Belkin M, Niyogi P (2003). Laplacian eigenmaps for dimensionality reduction
and data representation. \textit{Neural Computation,} 15 (6), 1373-1396

Benz\'{e}cri JP (1973a)\ \textit{L'Analyse des Donn\'{e}es: Vol. 1: La
Taxinomie}. Paris: Dunod

Benz\'{e}cri JP (1973b)\ \textit{L'Analyse des Donn\'{e}es: Vol. 2: L'Analyse
des Correspondances}. Paris: Dunod

Bolger DT, Alberts AC, Sauvajot RM, Potenza P, McCalvin C, Tran D, Mazzoni S,
Soule M (1997) Response of rodents to habitat fragmentation on coastal
southern California. \textit{Ecological Applications}, 7, 552--563

Brualdi RA, Parter SV, Schneider H (966) The diagonal equivalence of a
nonnegative matrix to a stochastic matrix. \textit{Journal of Mathematical
Analysis and Applications,} 16, 31--50

Choulakian V (2006) Taxicab correspondence analysis. \textit{Psychometrika,}
71, 333-345

Choulakian V (2016) Matrix factorizations based on induced norms.
\textit{Statistics, Optimization and Information Computing}, 4, 1-14

Choulakian V (2017) Taxicab correspondence analysis of sparse contingency
tables. \textit{Italian Journal of Applied Statistics,} 29 (2-3), 153-179

Choulakian V (2023) Notes on correspondence analysis of power transformed data
sets. https://arxiv.org/pdf/2301.01364.pdf

Choulakian V, Allard J, Mahdi S (2023) \ Taxicab correspondence analysis and
Taxicab logratio analysis: A comparison on contingency tables and
compositional data. \textit{Austrian Journal of Statistics,} 52, 39 -- 70

Choulakian V, Mahdi S (2024) Correspondence analysis with prespecified
marginals and Goodman's marginal-free correspondence analysis. In Beh,
Lombardo, Clavel (eds): Analysis of Categorical Data from Historical
Perspectives: Essays in Honour of Shizuhiko Nishisato, pp 335--352, Springer

Cuadras CM, Cuadras D, Greenacre M (2006) A comparison of different methods
for representing categorical data. \textit{Communications in Statistics -
Simulation and Computation}, 35(2), 447-459

Donoho DL, Grimes C (2003). Hessian eigenmaps: Locally linear embedding
techniques for high-dimensional data. \textit{Proceedings of the National
Academy of Sciences,} 100 (10), 5591-5596

Gifi, A. (1990). Nonlinear Multivariate Analysis. Wiley, New York

Goodman LA (1991) Measures, models, and graphical displays in the analysis of
cross-classified data. \textit{Journal of the American Statistical
Association}, 86 (4), 1085-1111

Goodman LA (1996) A single general method for the analysis of cross-classified
data: Reconciliation and synthesis of some methods of Pearson, Yule, and
Fisher, and also some methods of correspondence analysis and association
analysis.\textit{\ Journal of the American Statistical Association}, 91, 408-428

Graeve M, Greenacre M (2020) The selection and analysis of fatty acid ratios:
A new approachfor the univariate and multivariate analysis of fatty acid
trophicmarkers in marine pelagic organisms. \textit{Limnology and
Oceanography: Methods}, 18(5), 183-195

Greenacre M (1984) \textit{Theory and Applications of Correspondence
Analysis}. Academic Press

Greenacre M (2009) Power transformations in correspondence analysis.
\textit{Computational Statistics \& Data Analysis,} 53(8), 3107-3116

Greenacre M (2010) Log-ratio analysis is a limiting case of correspondence
analysis. \textit{Mathematical Geosciences,} 42, 129-134

Greenacre M (2020) \textit{R package easyCODA. }Available at\textit{:}

https ://CRAN.R-project.org/package=\textit{easyCODA}

Greenacre M (2024). The chiPower transformation: a valid alternative to
logratio transformations in compositional data analysis. \textit{Adv Data Anal
Classif} 18, 769--796. https://doi.org/10.1007/s11634-024-00600-x

Greenacre M, Lewi P (2009) Distributional equivalence and subcompositional
coherence in the analysis of compositional data, contingency tables and
ratio-scale measurements. \textit{Journal of Classification}, 26, 29-54

Greenacre M, Nenadic O, Friendly M (2022) \textit{R Packa ca. }Available
at\textit{:}

https ://CRAN.R-project.org/package=ca

Hotelling H (1933) Analysis of a complex of statistical variables into
principal components. \textit{Journal of Educational Psychology}, 24(6), 417-441

Idel M (2016) A review of matrix scaling and Sinkhorn's normal form for
matrices and positive maps. https://arxiv.org/pdf/1609.06349.pdf

Kruskal JB (1964) Nonmetric multidimensional scaling: A numerical method.
\textit{Psychometrika,} 29 (2), 115-129

Landa B, Zhang T, Kluger Y (2022) Biwhitening reveals the rank of a count
matrix. \textit{SIAM J Math Data Sci}, 4(4):1420-1446. Also available at: https://arxiv.org/pdf/2103.13840.pdf

Le Roux B, Rouanet H (2004) \textit{Geometric Data Analysis}. Springer NY

Lewi PJ (1976) Spectral mapping, a technique for classifying biological
activity profiles of chemical compounds. \textit{Arzneim Forsch (Drug Res),} 26:1295--1300

Loukaki M (2023) Doubly stochastic arrays with small support.
\textit{Australasian Journal Of Combinatorics}, 86(1) 136--148

Ma R, Sun E, Zou J (2023) A spectral method for assessing and combining
multiple data visualizations.\ Nat Commun 14, 780
https://doi.org/10.1038/s41467-023-36492-2. Available also at: https://arxiv.org/pdf/2210.13711

Madre JL (1980) M\'{e}thodes d'ajustement d'un tableau \`{a} des marges.
Cahiers de l'analyse des donn\'{e}es 5(1) 87-99, http://eudml.org/doc/87981%
$>$%
.Madre (1980)

Marshall AW, Olkin I, Arnold BC (2009) \textit{Inequalities: Theory of
Majorization and Its Applications}. Springer NY

McInnes L, Healy J, Melville J (2018). Umap: Uniform manifold approximation
and projection for dimension reduction. arXiv:1802.03426

Moon KR, van Dijk D, Wang Z, Gigante S, Burkhardt DB, Chen WS, Yim K, van den
Elzen A, Hirn MJ, Coifman RR, Ivanova NB, Wolf G, Krishnaswamy S (2019).
Visualizing structure and transitions in highdimensional biological data.
\textit{Nature Biotechnology} 37 (12), 1482-1492

Murtagh F (2005) \textit{Correspondence Analysis And Data Coding With Java And
R}. Taylor \& Francis, UK

Quinn G, Keough M (2002) \textit{Experimental Design and Data Analysis for
Biologists}. Cambridge University Press, Cambridge, UK

Rao CR (1973) \textit{Linear Statistical Inference and Its Applications}. John
Wiley and Sons: NY

Roweis ST, Saul LK (2000) Nonlinear dimensionality reduction by locally linear
embedding. \textit{Science,} 290 (5500), 2323-2326

Sammon JW (1969) A nonlinear mapping for data structure analysis. \textit{IEEE
Transactions on Computers,} 100 (5), 401-409

Scholkopf B, Smola A, Muller KR (1997) Kernel principal component analysis. In
\textit{International Conference on Artificial Neural Networks}, 583-588, Springer

Sinkhorn R (1964) A relationship between arbitrary positive matrices and
doubly stochastic matrices.\textit{ Ann. Math. Statist.} 35, 876-879

Sinkhorn R, Knopp P (967) Concerning nonnegative matrices and doubly
stochastic matrices. Pacific J. Math., 21(2):343--348, 1967. URL:
https://projecteuclid.org: 443/euclid.pjm/1102992505

Tenenbaum JB, Silva V, Langford JC (2000) A global geometric framework for
nonlinear dimensionality reduction. \textit{Science,} 290 (5500), 2319-2323

Torgerson W (1952) Multidimensional scaling: I. theory and method.
\textit{Psychometrika}, 17(4), 401--419

Tsagris M, Preston S, Wood A (2011) A data-based power transformation for
compositional data. In \textit{4th International Workshop on Compositional
Data Analysis}, J.J. Egozcue, R. Tolosana-Delgado, and M.I. Ortego, eds.,
Girona, Spain. International Center for Numerical Methods in Engineering
(CIMNE) Barcelona, Spain, 2011, pp. 1--8. Available at http://hdl.handle.net/2117/366618

Tsagris M, Alenazi A, Stewart C (2023) Flexible non-parametric regression
models for compositional response data with zeros. \textit{Statistics}
\textit{and Computing}, 33, 1-17. Available at
https://doi.org/10.1007/s11222-023-10277-5, 106

Tukey JW (1977) \textit{Exploratory Data Analysis}. Addison-Wesley: Reading, Massachusetts

Yule GU (1912) On the methods of measuring association between two attributes.
\textit{JRSS}, 75, 579-642

van der Maaten L, Hinton G (2008) Visualizing data using t-SNE.
\textit{Journal of Machine Learning Research} 9(Nov), 2579-2605

Ward K, Macfarlane G (020) \textit{Package ipfr in R. }Available at:

\bigskip https ://CRAN.R-project.org/package=ipfr

\bigskip

\bigskip

\bigskip\textbf{Appendix}

\bigskip iteration = 1500 or 1499 \ \ \ \ ca(dij)\$sv

[1] 1.0000000 1.0000000 1.0000000 1.0000000 1.0000000 1.0000000 1.0000000 1.0000000

[9] 1.0000000 1.0000000 1.0000000 1.0000000 1.0000000 1.0000000 1.0000000 1.0000000

[17] 1.0000000 1.0000000 1.0000000 1.0000000 1.0000000 1.0000000 1.0000000 1.0000000

[25] 1.0000000 1.0000000 1.0000000 1.0000000 1.0000000 1.0000000 1.0000000 1.0000000

[33] 1.0000000 1.0000000 1.0000000 1.0000000 1.0000000 1.0000000 1.0000000 1.0000000

[41] 1.0000000 1.0000000 1.0000000 1.0000000 1.0000000 1.0000000 1.0000000 1.0000000

[49] 1.0000000 1.0000000 1.0000000 1.0000000 1.0000000 1.0000000 1.0000000 1.0000000

[57] 1.0000000 1.0000000 1.0000000 1.0000000 1.0000000 1.0000000 1.0000000 1.0000000

[65] 1.0000000 1.0000000 1.0000000 1.0000000 1.0000000 1.0000000 1.0000000 1.0000000

[73] 1.0000000 1.0000000 1.0000000 1.0000000 1.0000000 1.0000000 1.0000000 1.0000000

[81] 1.0000000 1.0000000 1.0000000 1.0000000 1.0000000 1.0000000 1.0000000 1.0000000

[89] 1.0000000 1.0000000 1.0000000 1.0000000 1.0000000 1.0000000 1.0000000 1.0000000

[97] 1.0000000 1.0000000 1.0000000 1.0000000 1.0000000 1.0000000 1.0000000 1.0000000

[105] 1.0000000 1.0000000 1.0000000 1.0000000 1.0000000 1.0000000 0.9999999 0.9999999

[113] 0.9999998 0.9999998 0.9999998 0.9999996 0.9999994 0.9999994 0.9999987 0.9999982

[121] 0.9999976 0.9999973 0.9999971 0.9999949 0.9999940 0.9999933 0.9999916 0.9999907

[129] 0.9999879 0.9999860 0.9999857 0.9999830 0.9999829 0.9999744 0.9999565 0.9999042

[137] 0.9998845 0.9998704 0.9998388 0.9998303 0.9998243 0.9998098 0.9998003 0.9997900

[145] 0.9997572 0.9997502 0.9997118 0.9997023 0.9997001 0.9996577 0.9995629 0.9995457

[153] 0.9995050 0.9994306 0.9993534 0.9993510 0.9993496 0.9993461 0.9993224 0.9993088

[161] 0.9993075 0.9992959 0.9992790 0.9992493 0.9992439 0.9992186 0.9992142 0.9992085

[169] 0.9992076 0.9991986 0.9991561 0.9990049 0.9989804 0.9989786 0.9989749 0.9989159

[177] 0.9989024 0.9987829 0.9987727 0.9987690 0.9987292 0.9986991 0.9984546 0.9984065

[185] 0.9984009 0.9983052 0.9981344 0.9981019 0.9980984 0.9980983 0.9980219 0.9979210

[193] 0.9978944 0.9978376 0.9978274 0.9977364 0.9977316 0.9977285 0.9977223 0.9976919

[201] 0.9976404 0.9976010 0.9975551 0.9975516 0.9974197 0.9973882 0.9973433 0.9973228

[209] 0.9971855 0.9969870 0.9969393 0.9968935 0.9968566 0.9967788 0.9967049 0.9966357

[217] 0.9964829 0.9964476 0.9963342 0.9962191 0.9961630 0.9961446 0.9960815 0.9958977

[225] 0.9958652 0.9958554 0.9957947 0.9957758 0.9957730 0.9957716 0.9957323 0.9956531

[233] 0.9955773 0.9952718 0.9952445 0.9951317 0.9951046 0.9949589 0.9948421 0.9948358

[241] 0.9946760 0.9946431 0.9943875 0.9942713 0.9941642 0.9940914 0.9940410 0.9940356

[249] 0.9939414 0.9937308 0.9935095 0.9933060 0.9932102 0.9931092 0.9930615 0.9929306

[257] 0.9928601 0.9928295 0.9925974 0.9925626 0.9924373 0.9922767 0.9922610 0.9922431

[265] 0.9920729 0.9920425 0.9919547 0.9918366 0.9918350 0.9917942 0.9915453 0.9914892

[273] 0.9911905 0.9911121 0.9909367 0.9908857 0.9907796 0.9906793 0.9906354 0.9906282

[281] 0.9905022 0.9904279 0.9904134 0.9903917 0.9903752 0.9903060 0.9900805 0.9899054

[289] 0.9898829 0.9897441 0.9897386 0.9895356 0.9895161 0.9893045 0.9892339 0.9891573

[297] 0.9890011 0.9889965 0.9888289 0.9888264 0.9887329 0.9885884 0.9884499 0.9884249

[305] 0.9883281 0.9880578 0.9879792 0.9878656 0.9877138 0.9876940 0.9875869 0.9873000

[313] 0.9872883 0.9872604 0.9871252 0.9870861 0.9867917 0.9867712 0.9867281 0.9863939

[321] 0.9861168 0.9860737 0.9859532 0.9857569 0.9857475 0.9857252 0.9856825 0.9854015

[329] 0.9852575 0.9851506 0.9849342 0.9847880 0.9846818 0.9846278 0.9845084 0.9844634

[337] 0.9843424 0.9842373 0.9835844 0.9833021 0.9832924 0.9829480 0.9829328 0.9829049

[345] 0.9828540 0.9826716 0.9825560 0.9825123 0.9824157 0.9822316 0.9821151 0.9820461

[353] 0.9819228 0.9818451 0.9813221 0.9813052 0.9810376 0.9810045 0.9809847 0.9805979

[361] 0.9805581 0.9803791 0.9800348 0.9797959 0.9792255 0.9789450 0.9789224 0.9788074

[369] 0.9785766 0.9784028 0.9783570 0.9781365 0.9779832 0.9779813 0.9779671 0.9779209

[377] 0.9778849 0.9775773 0.9774715 0.9773847 0.9772981 0.9772069 0.9771841 0.9770516

[385] 0.9767973 0.9767793 0.9767718 0.9763481 0.9762838 0.9755859 0.9755749 0.9755571

[393] 0.9753048 0.9753038 0.9751698 0.9751271 0.9749917 0.9745571 0.9745226 0.9744746

[401] 0.9743928 0.9743431 0.9742835 0.9741217 0.9738488 0.9735054 0.9733233 0.9727307

[409] 0.9726781 0.9724381 0.9723704 0.9723146 0.9720727 0.9718992 0.9712859 0.9712407

[417] 0.9711802 0.9705660 0.9705552 0.9704082 0.9702667 0.9698933 0.9698900 0.9692294

[425] 0.9688700 0.9684887 0.9682634 0.9682319 0.9679514 0.9676946 0.9675131 0.9669772

[433] 0.9668160 0.9667117 0.9666040 0.9663994 0.9662453 0.9660918 0.9659918 0.9657493

[441] 0.9656911 0.9656375 0.9649920 0.9644599 0.9644458 0.9643957 0.9642596 0.9642332

[449] 0.9640634 0.9636526 0.9636256 0.9633100 0.9631252 0.9631247 0.9627706 0.9627315

[457] 0.9625069 0.9624609 0.9622236 0.9617261 0.9616493 0.9609082 0.9608423 0.9607209

[465] 0.9606808 0.9602031 0.9601826 0.9598969 0.9595925 0.9592491 0.9590916 0.9587804

[473] 0.9583899 0.9580232 0.9577250 0.9567925 0.9564024 0.9561637 0.9560063 0.9560031

[481] 0.9559409 0.9555687 0.9550438 0.9548637 0.9548146 0.9547520 0.9533340 0.9530938

[489] 0.9528115 0.9527882 0.9517189 0.9511896 0.9508099 0.9506287 0.9502582 0.9494205

[497] 0.9494071 0.9487999 0.9487477 0.9486811 0.9474266 0.9471492 0.9466498 0.9464657

[505] 0.9463625 0.9460308 0.9460213 0.9459221 0.9455041 0.9449110 0.9439099 0.9435329

[513] 0.9434508 0.9434305 0.9432871 0.9428326 0.9426490 0.9420296 0.9412092 0.9411211

[521] 0.9394350 0.9388870 0.9380430 0.9376750 0.9372849 0.9366966 0.9358015 0.9357109

[529] 0.9350730 0.9345864 0.9337436 0.9330633 0.9299461 0.9295742 0.9293285 0.9290859

[537] 0.9280448 0.9278056 0.9269726 0.9257694 0.9257612 0.9251830 0.9244718 0.9240081

[545] 0.9239780 0.9239617 0.9215778 0.9215532 0.9205492 0.9205181 0.9194363 0.9161280

[553] 0.9149813 0.9148949 0.9139334 0.9105601 0.9095867 0.9068417 0.9055099 0.9014791

[561] 0.9006996 0.8992565 0.8975730 0.8956672 0.8938989 0.8914348 0.8865637 0.8766786

[569] 0.8766368 0.8722169 0.8708924 0.8683782 0.8653681 0.8551611 0.8551513 0.8411667

[577] 0.8364237 0.8306574 0.8167284 0.8102990 0.8078165 0.7959309 0.7813042 0.7049943

[585] 0.6749376 0.5941499 0.4983027 0.3236741
\end{document}